\documentclass[12pt]{article}
\usepackage{authblk}
\usepackage{amsmath, amssymb}
\usepackage{graphicx}
\usepackage{hyperref}
\usepackage{graphicx}
\usepackage{float}
\usepackage{doi}
\usepackage{subcaption}
\usepackage{xcolor}
\usepackage{siunitx}
\usepackage{xcolor}
\usepackage{colortbl}
\usepackage{booktabs}
\usepackage[a4paper, margin=1in]{geometry}
\newcommand{\alphafit}{\num{0.2506}}
\newcommand{\etafit}{\num{1.0924}}
\newcommand{\Hnaughtfit}{\num{64.217}}
\newcommand{\Mfit}{\num{-19.539}}
\newcommand{\chiSqMin}{\num{1038.659}}
\newcommand{\DoF}{\num{1044}}
\newcommand{\chiSqRed}{\num{0.995}}

\title{\textbf{Non-Singular Bouncing cosmology from Phantom Scalar–Gauss-Bonnet Coupling: Reconstruction with Observational Insights}}
\author[1]{Khandro K. Chokyi\thanks{Email: \texttt{khandro.chokyi1@s.amity.edu}; \texttt{khankalmaths2023@outlook.com}}}
\author[1]{Surajit Chattopadhyay\thanks{Corresponding author. Email: \texttt{surajitchatto@outlook.com}; \texttt{schattopadhyay1@kol.amity.edu}}}

\affil[1]{Department of Mathematics, Amity University Kolkata, Major Arterial Road, Action Area II, Kolkata 700135, India}

\date{\today}

\begin{document}

\maketitle

\begin{abstract}
We examine non-singular bounce cosmology within the framework of a phantom scalar field coupled to the Gauss-Bonnet term in both non-viscous and bulk-viscous cases. Using the scale factor ansatz $\alpha(t)=\left(\frac{\alpha}{\eta}+t^2\right)^{\frac{1}{2 \eta}}$, we reconstruct the scalar field potential $V(t)$, and observe a smooth potential well centered at the bounce point. The resulting energy density, pressure, and equation-of-state parameter show NEC violation necessary for successful bounce, while viscosity controls post-bounce dynamics with a positive and smooth squared speed of sound. In contrast, for the non-viscous model, sharp divergences occur just \textcolor{black}{at the bounce and continues to be negative in the expanding phase}, which in turn emphasises the stabilising role of dissipative effects. The energy condition analysis indicates a temporary NEC and SEC violation in the viscous scenario, whereas its persistent violation within the non-viscous model suggests a continuous accelerated expansion. Observational viability is found through Bayesian MCMC fitting in regards to the Pantheon+ supernova data, with best-fit parameters providing a reduced chi-squared of $\chi_{red}^2 =0.995$ while the inflation observables derived from the reconstructed potential place our model predictions inside $68\%$ CL Planck 2018 confidence contours. Our findings suggest that bounce cosmologies could offer a physically reasonable and observationally acceptable alternative or pre-inflationary scenario, while highlighting the role that viscosity could play for a stable and smooth cosmological evolution.
\end{abstract}

\section{Introduction}
In recent years, cosmology has been characterized as entering an era of precision comparable to that of particle physics \cite{guth2002inflation,mukhanov2005physical, peter2009primordial}. Although this is perhaps exaggerated slightly, the claim is not without basis. Observations at high precision, especially from the PLANCK satellite, have allowed scientists not only to compare various theoretical models but also to prefer certain mechanisms in the context of the early epochs of the universe. Consequently, inflationary cosmology \cite{guth2023inflationary,linde2005particle,lazarides2002inflationary} has become the front-runner, and probably the most natural, explanation for these observations. Popularity, however, does not make inflation the absolute truth, and it demands extensive testing to eliminate all other alternatives. The initial singularities \cite{borde1996singularities} encountered in inflationary cosmology motivated academics to widen their approach towards other possible explanations. Presently, the only promising alternative receiving robust attention includes nonsingular bouncing models wherein the scale factor $a(t)$ never reaches zero at $t=0$. Instead, the cosmos expands when $t\geq0$ and contracts during $-\infty<t\leq0$, i.e. a contracting phase precedes the expanding phase without any initial singularity. Ijjas and Steinhardt \cite{ijjas2018bouncing} have introduced the "wedge diagram" that helps visualise the cosmological evolution in terms of cosmological bounce. The cosmic singularity is substituted by a bounce where the scale factor decreases to a
finite critical size, much greater than the Planck length, and then bounces. The non-singular(classical) bouncing cosmology claims to solve the cosmic singularity problem \cite{heller2000cosmological} and the horizon problem\cite{bolotin2013thousand} while also explaining the smoothness and flatness of the universe \cite{helbig2021arguments}. For an extensive review of the various classical bouncing models, \cite{battefeld2015critical}
can be looked into as it provides the reader with a critical assessment of the proliferating bouncing models present in literature, and can serve as a guide for future research. In a similar vein, one can also refer to \cite{cai2014exploring} where the author has reviewed cosmological bounce theory considering cosmological surveys.\\
In the case of a non-singular bounce:
\begin{itemize}
    \item the scale factor decreases with time ($\dot{a}<0$), resulting in a contracting phase with a negative Hubble parameter ($H=\frac{\dot{a}}{a}<0$).
    \item  At the bouncing epoch, the scale factor contracts to a non-zero, finite critical size, causing the Hubble parameter to vanish ($H_{t=0}=0$).
    \item Positive acceleration causes the scale factor to rise with time, resulting in a positive Hubble parameter ($H=\frac{\dot{a}}{a}>0$)
\end{itemize}
Einstein's field equations for flat FLRW space are written as:
\begin{equation}\label{i1}
 H^2=\frac{1}{3}\rho   
\end{equation}
\begin{equation}\label{i2}
    \dot{H}=-\frac{1}{2}(p+\rho)
\end{equation}
From the requirements mentioned previously and Eq. (\ref{i2}), one observes that for bounce:
\begin{equation}
    [\rho+p]_{t=0}<0
\end{equation}
This means that at the bouncing point, the null energy condition(NEC) must be violated. It was implied in \cite{peter2001has} that in a cosmological setting, the violation of NEC calls for exotic matter. Bounce cosmology also provides a way to describe the NEC breach in the early universe \cite{rubakov2014null}. The Hawking-Penrose theorems \cite{hawking1970singularities} hold that an initial cosmic singularity is inevitable under the framework of GR in a homogeneous and isotropic scenario.  This depends on the minimal interaction of matter with gravity and validating NEC \cite{peter2002primordial}. Thus, among myriad approaches adopted, modifications to general relativity(GR) theory have also been investigated to account for the bouncing scenario. Incorporating matter that violates the NEC while including the theory of Einstein's GR may be possible, yet avoiding gradient and ghost instabilities \cite{cline2004phantom} is a major challenge one has to face while adopting this route. Therefore, it may be more enticing to study cosmological bounce within the modified gravity framework \cite{shankaranarayanan2022modified}. 
\par
Among the various choices of modified gravity available in literature(teleparallel gravity \cite{aldrovandi2012teleparallel,cai2016f}, $f(R)$ gravity \cite{capozziello2008cosmography,sotiriou2010f}, braneworld gravity \cite{maartens2010brane}, Horndeski gravity \cite{kobayashi2019horndeski}, scalar-tensor theory \cite{fujii2003scalar} and more), phantom scalar field \cite{johri2004phantom}—a scalar field with a negative kinetic term—has shown effectiveness when it comes to NEC violation without necessarily depending on exotic matter components. Adding geometric corrections to GR may further enhance the viability and stability of the bouncing models. In this context, the Gauss-Bonnet(GB) term is a promising example. The gravitational theory known as Einstein-Gauss-Bonnet theory \cite{zwiebach1985curvature,gross1987quartic,dadhich2010characterization, fernandes20224d} introduces the Gauss-Bonnet scalar within the gravitational action integral, and it belongs to the family of Lovelock’s theories of gravity \cite{lovelock1971einstein}. Importantly, the Gauss-Bonnet term produces second-order field equations, avoiding the Ostrogradsky instability \cite{woodard2015theorem}. It is the primary higher-curvature correction to the Einstein-Hilbert action. Although this factor by itself does not affect the gravitational field equations in four-dimensional spacetime, its contribution becomes important when a coupling with a scalar field is introduced, and this scalar-Gauss-Bonnet coupling results in non-trivial dynamics. It should also be noted that in a four-dimensional manifold, the Gauss-Bonnet scalar is a topological invariant, thus adding a boundary term to the gravitational action integral. However, in five- or higher-dimensional manifolds, the Gauss-Bonnet scalar brings new geometrodynamical elements into the field equations, resulting in new phenomena. The GB term no longer remains topologically invariant when multiplied by the coupling function. This adds extra degrees of freedom to the field equations while also sustaining the second-order nature of gravitational theory. Studies on such non-zero coupling with a scalar field can be found in \cite{koivisto2007cosmology,sami2005fate}. This field has also been extensively used to model the inflationary paradigm \cite{fomin2025scalar,odintsov2018viable,supriyadi2022inflation,chakraborty2018inflation}. Other notable works in this context can be found in \cite{oikonomou2024einstein,gohain2024emergent,odintsov2020non,rashidi2020gauss,motaharfar2016warm}. The GB term is also reported to have affected the dynamics in the context of quintessence scalar field such that the equation of state(EoS) parameter crosses the phantom divide($\omega=-1$) \cite{tsujikawa2007string}. At this point, we would like to bring to the reader's attention a study undertaken by Papagiannopoulos \textit{et.al.} \cite{papagiannopoulos2025avoiding}. This relatively new work delves into the dynamical analysis of the phantom scalar field with the Gauss-Bonnet term. Furthermore, a chameleon mechanism \cite{khoury2004chameleon,khoury2004chameleonc} is enabled by assuming a coupling between the matter source and the scalar field. The authors conclude that the EoS parameter shows crossing of the phantom divide while avoiding Big Rip singularity due to the interaction between the matter and the field. In a recent study, Rani et al. \cite{Rani2025} examined nonsingular bouncing models within the framework of Aether scalar–tensor gravity. They showed that the inclusion of the Aether field enables a smooth transition from contraction to expansion, thereby resolving the initial singularity problem and offering a consistent alternative to inflationary scenarios. Bouncing cosmology in Gauss-Bonnet gravity has been explored in \cite{bamba2014bouncing,bamba2015bounce,odintsov2020unification,oikonomou2015singular,haro2015bounce,ilyas2023probing,singh2022bouncing,makarenko2018generalized,maeda2007braneworld,ganiou2022reconstruction} and the cosmological dynamics of phantom and quintessence scalar fields have also been investigated(with or without Gauss-Bonnet coupling) in several studies, we observe, to the best of our knowledge, that bounce cosmology has not been explored in the context of the combination of phantom scalar field coupled with GB term and matter as established in \cite{papagiannopoulos2025avoiding}. This constitutes a probable literature gap. Our study intends to address this issue by examining two distinct bouncing models: one without viscosity and the other with bulk viscosity, within the framework of the aforementioned phantom scalar field with the Gauss-Bonnet term.

We have referenced two keystone datasets: \textit{Planck 2018} \cite{Akrami2020PlanckInflation} and \textit{Pantheon+SH0ES} \cite{Scolnic2021PantheonFull} to contend our model parameters with the current observational data available. The 2018 release of the Planck cosmic microwave background (CMB) anisotropy measurements was proven to be consistent with the previously released data. In addition to this, it also reduced uncertainties by providing better characterizations of polarization across multipoles. The \textit{Pantheon+SH0ES}(see \cite{brout2022pantheon+, riess2022comprehensive} ) compilation, on the other hand, offers light curve measurements of 1550 SN objects, spanning a wide redshift range and incorporating low-redshift calibrators for improved distance-ladder accuracy. This dataset is a compilation of all SNe Ia discovered for $z<<0.01$.
In our work, these datasets serve to augment the reconstructed bounce dynamics within the Gauss--Bonnet phantom scalar field framework against empirical results, and thus complement the theoretical foundation in our work. Specifically, we have used the supernova-derived distance moduli to evaluate late-time expansion consistency to justify our choice of the scale factor and to validate the model in the $n_s$--$r$ plane. Even though the conceptual and analytical structure of the bouncing scenario continues to be the major focus of our work, this data-driven validation warrants the observational viability without shifting much focus towards observational analysis.

In view of the survey of literature presented above, we iterate once again that though inflation is widely accepted, it does not completely resolve the problem of the initial singularity and certain related instabilities. This has motivated the study of bouncing cosmology as an attractive alternative, where the universe first contracts and then expands smoothly without hitting a singularity. Such models can naturally address the horizon and flatness problems and can offer a consistent description of the early universe. In particular, a phantom scalar field coupled with the Gauss--Bonnet term provides a useful mechanism for realization of a non-singular bounce and also capable of maintaining stability. Moreover, the inclusion of viscosity may play an important role in regulating the dynamics after the bounce. With this motivation, in the present work we investigate the bouncing behaviour in a Gauss--Bonnet phantom scalar field framework, both with and without viscosity, and test its physical as well as observational viability. The structure of our paper is as follows:
In Section \ref{1}, we have laid out the basic equations governing the Gauss-Bonnet scalar-field cosmology. Following this, in Section \ref{2}, we have analysed the bouncing model-I within the gravitational theory specified in the previous section. Within the same framework, we have studied bouncing cosmology for the second model, which allows a bulk viscosity term in Section \ref{3}. The study on the energy conditions for both models has been elucidated in Section \ref{4}. The study on the squared speed of sound for stability analysis of both models is discussed in Section \ref{5}. Section \ref{6} presents an observational analysis of the reconstructed dynamics, employing data from Planck 2018 and Pantheon+SH0ES to assess the model's compatibility with current cosmological constraints. Finally, we present our conclusions in Section \ref{7}.
 
\section{Basic Formalism}\label{1}
In this Section, we have discussed the underlying equations of the gravitational dynamics and scalar field that we have applied to study the cosmological bouncing solution in the subsequent sections.
\begin{center}
    \textbf{Gravitational theory}
\end{center}
    For our work, we have considered the Einstein-Gauss-Bonnet scalar theory, which is defined by the action integral \cite{fomin2018cosmological}:
    \begin{equation}\label{g1}
        S=\int d^4 x\sqrt{-g}\left(R-f(\phi)G+\frac{1}{2}g^{\mu K}\bigtriangledown_{\mu}\phi \bigtriangledown_{K}\phi-V(\phi)-g(\phi)L_{m}(x^\nu)\right)
    \end{equation}
    where the Ricci scalar $R$ is for the four dimensional $g_{\mu\nu}$, $G$ is the Gauss-Bonnet scalar that is topologically invariant since $g$ is four-dimensional. The Lagrangian function for the matter source is represented by $L_{m}(x^\nu)$ and the scalar field $\phi$ is coupled to the matter source as well as to the Gauss-Bonnet scalar $G$. The energy transfer between $\phi$ and $L_{m}(x^\nu)$ is facilitated by the function $g(\phi)$. We note that the condition of constant $f(\phi)$ reduces the gravitational model in Eq. (\ref{g1}) to GR. The coupling between the scalar field and the Gauss-Bonnet scalar is characterized by the function $f(\phi)$.\\
    In the context of the scalar field, $\phi$, the field chosen is a phantom field, i.e. the kinetic energy term is negative. This in turn, hints at the violation of the weak energy condition(WEC) thus allowing for a negative energy density. The gravitational field equations are \cite{fomin2018cosmological}:
    \begin{equation}\label{g2}
        R_{\mu \nu}-\frac{1}{2}R g_{\mu\nu}=T^{G}_{\mu \nu}+T^{\phi}_{\mu \nu}+f(\phi)T^{m}_{\mu \nu}
    \end{equation}
    where $T^{m}_{\mu \nu}=\frac{\delta L_{m}}{\delta g_{\mu \nu}}$ is the energy-momentum tensor for the matter source, $T^{G}_{\mu \nu}$ is the effective energy-momentum tensor which assigns the Gauss-Bonnet scalar's geometrodynamical degrees of freedom, and $T^{\phi}_{\mu \nu}$ is the energy-momentum tensor for the phantom scalar field. 
    \begin{center}
        FLRW Universe
    \end{center}
We consider the field equations for an isotropic and spatially flat FLRW background geometry with line element:
\begin{equation}\label{F1}
 ds^2=-N^{2}(t)dt^2+a^2(t)(dx^2+dy^2+dz^2)   
\end{equation}
Here, $N(t)$ is a lapse function and $a(t)$ is the scale factor, which also describes the universe's radius. The expansion rate is calculated as $\theta=3H$, for a commoving observer $u^\mu=\delta_{t}^\mu$. The Hubble function $H=\frac{1}{N}\frac{\dot{a}}{a}$ and $\dot{a}=\frac{da}{dt}$. In addition to this, we are considering a pressureless, isotropic fluid that corresponds to the dark matter of the Universe. Hence, $L_{m}=\rho_{m_0} a^{-3}$. Without loss of generality, $N(t)=1$ can be assumed, and the field equations are \cite{millano2023global}:
\begin{equation}\label{F2}
  0=3H^2+\frac{1}{2}\dot{\phi}^2-V(\phi)\textcolor{black}{-}g(\phi)\rho_{m_0}a^{-3}-24f_{,\phi}H\dot{\phi}  
\end{equation}
\begin{equation}\label{F3}
 \textcolor{black}{0=2\dot{H}+3H^2-\Big(\frac{1}{2}\dot{\phi}^2+V(\phi)+16(H^2+\dot{H})Hf_{,\phi}\dot{\phi}+8H(\dot{\phi}^{2}f_{,\phi\phi}+f_{,\phi}\ddot{\phi})\Big)}
\end{equation}

\begin{equation}\label{F4}
  0=\ddot{\phi}+3H\dot{\phi}-V_{,\phi}-24H^2f_{,\phi}(H^2+\dot{H})-g_{,\phi}\rho_{m_0}a^{-3}  
\end{equation}
Equivalently, the above field equations can be written as:
\begin{equation}\label{F5}
    3H^2=\rho_{eff}
\end{equation}
\begin{equation}\label{F6}
    2\dot{H}+3H^2=-p_{eff}
\end{equation}
In this case, 
\begin{equation}\label{F7}
\rho_{eff}=-\frac{1}{2}\dot{\phi}^2+V(\phi)\textcolor{black}{+}g(\phi)\rho_{m_0}a^{-3}+24f_{,\phi}H^2\dot{\phi}    
\end{equation}
and
\begin{equation}\label{F8}
 \textcolor{black}{p_{eff}=-\Big(\frac{1}{2}\dot{\phi}^2+V(\phi)+16(H^2+\dot{H})Hf_{,\phi}\dot{\phi}+8H(\dot{\phi}^2f_{,\phi\phi}+f_{,\phi}\ddot{\phi})\Big)}   
\end{equation}
These formulas show how the dynamics of the scalar field and its couplings affect the universe's expansion history. \textcolor{black}{For detailed derivation of the field equations, please refer to Appendix \ref{a}}. Achieving a bouncing solution, which avoids the initial singularity and permits a transition from contraction to expansion, depends critically on the sign and behaviour of $f_{,\phi}$ \cite{nojiri2005modified}. Higher-derivative corrections are added to the typical cosmic dynamics when the Gauss-Bonnet term connected to the scalar field is present. Bouncing scenarios \cite{cai2012bounce}, in which the cosmos smoothly moves from contraction to expansion, avoiding the Big Bang singularity, can be supported by these corrections. A crucial prerequisite for achieving a nonsingular bounce is the null energy condition, which can be violated due to the scalar field's phantom nature (negative kinetic energy) \cite{bamba2014phantom}.
The theory of the bouncing scenario was proposed as an alternative to Big Bang cosmology. Bouncing models show a contracting phase before it leads to a non-singular bounce, which implies that the universe's expansion ($a(t)$) decreases with time as $\dot{a}(t)<0$. Thus, during the bouncing epoch, the Hubble parameter $H$ approaches zero and the deceleration parameter $q$ becomes singular. This means that for the bouncing scenario, the Hubble parameter's value transitions from a negative value to a positive value, and at the bouncing point($t=0$), $H=0$. As far as the scale factor is concerned, it increases during the expansion phase. 
For our model to show the bouncing properties, we have begun with a bouncing scale factor given by \cite{abdussattar2011role}:
\begin{equation}\label{b1}
   a=\left(\frac{\alpha}{\eta}+t^2\right)^{\frac{1}{2\eta}} 
\end{equation}
where $\eta$ and $\alpha$ are parameters that are suitably chosen to provide a bouncing behaviour.

The scale factor exhibits features that indicate a nonsingular, bouncing cosmology. Contrary to standard Big Bang scenarios, where $a(t) \to 0$ as $t \to 0$, this form ensures a minimum nonzero value of the scale factor at $t=0$:
\begin{equation}
a_{\text{min}} = \left(\frac{\alpha}{\eta}\right)^{\frac{1}{2\eta}}.
\end{equation}
This property does not allow a curvature singularity but does not prevent the universe from having a smooth transition from pre-bounce contraction ($t<0$) to post-bounce expansion ($t>0$), which is the characteristic of a cosmological bounce. We consider the neighbourhood of $t = 0$. In this neighbourhood, we approximate:
\begin{equation}
a(t) \approx \left(\frac{\alpha}{\eta}\right)^{\frac{1}{2\eta}} \left(1 + \frac{\eta t^2}{\alpha}\right)^{\frac{1}{2\eta}} \approx \left(\frac{\alpha}{\eta}\right)^{\frac{1}{2\eta}} \left[1 + \frac{t^2}{2\alpha} + \mathcal{O}(t^4)\right].
\end{equation}
The above expression shows that the scale factor reaches a nonzero minimum value and increases in a quadratic pattern with time near the bounce. This confirms the non-singular nature of the cosmology.\\
\textcolor{black}{There are several other forms of bouncing scale factors that exist in literature \cite{pasqua2013new, molchanov2023cosmological, oliveira2014cosmological}, each one trying to highlight different physical details. Some of the standard examples include exponential bounces \cite{gul2024comprehensive,tripathy2019bouncing, bari2018cosmological} like $a(t) = a_0^{\lambda t^2}$, power-law bounces \cite{gielen2015perfect, debnath2021bouncing, sharif2025analysis} and then the matter-bounce-inspired versions \cite{cai2011matter, samaddar2025matter, brandenberger2009matter}, where the contracting phase mimics a dust-dominated universe. Exponential bounces keep the Hubble parameter and its derivatives regular, but they don’t really connect well to familiar cosmological epochs at late times. Power-law bounces, in contrast, are more flexible and let you tweak the bounce’s shape, but they add more free parameters, which makes reconstruction more complicated.
The scale factor we have chosen and it brings together several desirable features. Firstly, the scale factor never reaches zero at the bounce. Secondly, it admits simple analytic forms for $H(t)$ and $\dot{H}(t)$. This ensures a smooth flow from a contracting phase, through a non-singular bounce, and then shifts into a power-law expansion at late times. The parameter $\eta$ is especially useful as it controls the post-bounce expansion rate directly. Depending on its value, you can model either a matter-dominated or an accelerated expansion, which makes it suitable for late-time observational analysis. Thus, we have chosen this form because it’s analytically simple, uses as few parameters as possible, and links the early bounce with the universe’s later evolution in one consistent background.}

\begin{figure}[!ht]
    \centering
    \begin{subfigure}[b]{0.35\textwidth}
        \centering
\includegraphics[width=1.1\linewidth]{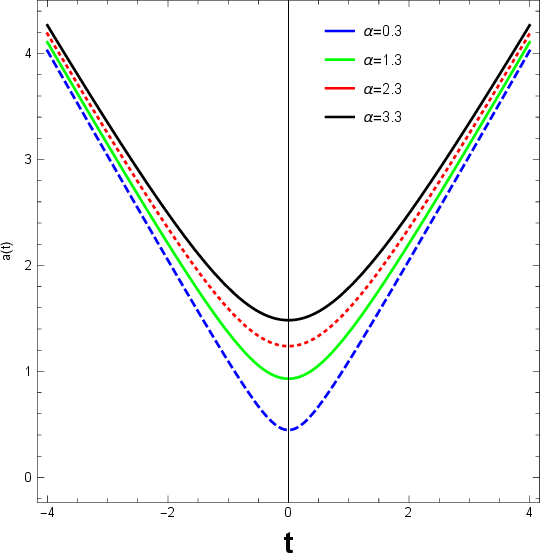}
        \caption{Evolution of $a(t)$.}
        \label{a1}
    \end{subfigure}
\hfill
    \begin{subfigure}[b]{0.35
\textwidth}
        \centering
\includegraphics[width=1.1\linewidth]{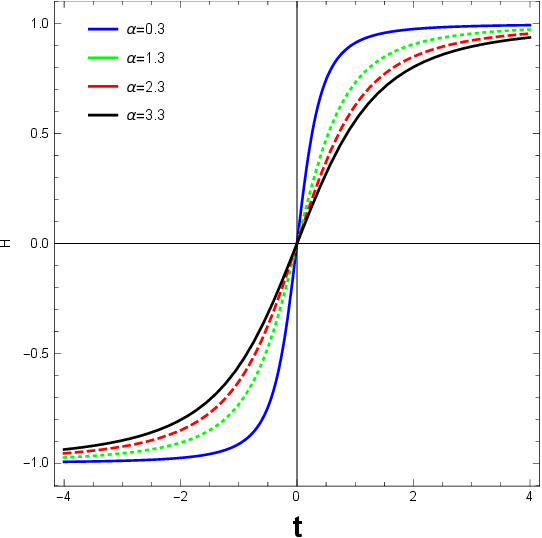}
        \caption{Evolution of $H$.}
        \label{h}
    \end{subfigure}
    \hfill
    \begin{subfigure}[b]{0.35
\textwidth}
        \centering
\includegraphics[width=1.1\linewidth]{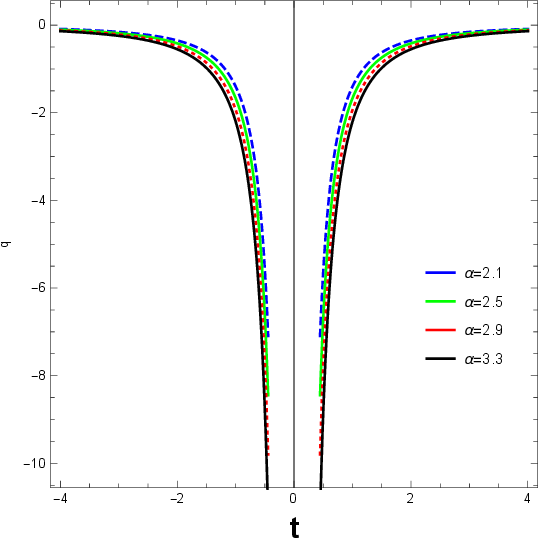}
        \caption{Evolution of $q(t)$.}
        \label{q}
    \end{subfigure}
    \caption{Evolution of $a$, $H$ and $q$ against cosmic time $t$ for different values of $\alpha$.}
    \label{f1}
    \end{figure}

The $H$ and its time derivative become:
\begin{align}
H(t) &= \frac{\dot{a}}{a} = \frac{t}{\left(\alpha + \eta t^2\right)}, \\
\dot{H}(t) &= \frac{\alpha - \eta t^2}{\left(\alpha+ \eta t^2\right)^2}.
\end{align}
At the point of bounce, we have $t=0$.  This leads to:
\begin{align}
H(0) = 0, \quad \dot{H}(0) = \frac{\eta}{\alpha} > 0,    
\end{align}
which confirms a bounce, since the Hubble parameter transitions from negative (contracting phase) to positive (expanding phase), and $\dot{H}(0) > 0$ satisfies the bouncing condition.
Additionally, the deceleration parameter is written in terms of the Hubble parameter as:
\begin{equation}\label{b2}
    q=-\frac{\dot{H}}{H^2}-1
\end{equation}
Using the equation for $H$ derived earlier, 
\begin{equation}
    q=-1-\frac{\alpha}{t^2}+\eta
\end{equation}
The behavior of the scale factor, the Hubble parameter, and the deceleration parameter($q$) with cosmic time are depicted in Fig. \ref{f1}. It is quite evident that the bouncing condition ($H<0$ at $t<0$, $H=0$ at $t=0$ and $H>0$ at $t>0$) is satisfied. We note some physical insights into the behaviour of the scale factor given. Let us consider the scenario where $|t|$ is considerably large so that $t^2$ dominates the right hand side and we get $a\propto t^{\frac{1}{\eta}}$. Now the behaviour of $\eta$ will tell us the further consequences as follows:
\begin{itemize}
    \item If we choose $\eta<1$, by some simple calculation we can show that $\ddot{a}>0$ which implies accelerated expansion.
    \item In case $\eta=1$, we get linear expansion.
    \item In case $\eta>1$, obviously $\ddot{a}<0$ indicating decelerated expansion
\end{itemize}
The behaviour illustrated above helps us to understand that the scale factor under consideration can interpolate between bounce at early stage and different possible expansion regimes at the later stages of the universe.

In Fig. \ref{a1}, we have pictorially depicted the bouncing scale factor. In Fig. \ref{h}, the behaviour of $H$ is explored. The evolving pattern of $H(t)$ around the bouncing point $t=0$ shows the symmetric and nonsingular behaviour. For the scale factor under consideration, in the pre-bounce regime ($t < 0$), $H(t)$ appears to be negative, indicating contraction, while in the post-bounce phase ($t > 0$), $H(t)$ appears to be positive, indicating expansion. At the bounce point ($t = 0$), $H(t)$ is vanishing, and that ensures a smooth transition from contraction to expansion. We further see in the figure that asymptotically, for large $|t|$, the Hubble parameter decays as \( H(t) \sim 1/(\eta t) \), characterising a quasi power-law behavior. Overall, the Hubble evolution substantiates a continuous and regular bouncing cosmology.
\textcolor{black}{We use the deceleration parameter $q(t)$ to get a clear sense of how the universe’s expansion changes during the bounce. We know that the scale factor $a(t)$ and the Hubble parameter $H(t)$ tell us there’s an occurrence of a non-singular bounce. But those don’t really show us when the universe shifts between speeding up and slowing down. That’s where $q(t)$ comes in. Its sign tells us right away if the expansion is accelerating or decelerating, so we can observe those transitions before and after the bounce. Something interesting happens at the bounce itself: $q(t)$ actually diverges. That mirrors the fact that H vanishes there, which is what you’d expect in any non-singular bouncing cosmology. After the bounce, the behaviour of $q(t)$ at late times tells us if the expansion keeps speeding up or starts to slow down, depending on the value of $\eta$, linking the early time bounce dynamics to the late-time cosmological dynamics. So, $q(t)$ doesn’t just function as an analytic element — it’s a key tool for diagnosing and understanding how the model evolves, from the bounce all the way to late-time expansion.}
\textcolor{black}{At this point in our study, we would like to emphasise that the use of a well-known bouncing scale factor as given in \eqref{b1}, does not trivialise the cosmological dynamics obtained in our work because even though it is already established in literature that the particular form of $a(t)$ gives a non-singular bounce at $t=0$, nonetheless, the evolution of the physical parameters like energy density, the effective pressure, the equation of state parameter, and the energy conditions are greatly influenced by the gravitational theory and the matter content that we have chosen. The current scenario, where the phantom scalar field connected to the Gauss–Bonnet term is being reconstructed, and considerations of viscous and non-viscous cases are combined leads to significant changes in the evolution after the bounce. In our present work, the key aspect is the reconstruction of the potential associated with the phantom-scalar field which, together with the viscous and non-viscous models under consideration, leads to non-trivialisation of the bounce dynamics.  Furthermore, the behaviour of the energy conditions, stability and slow-roll parameters are not determined by just the choice of the scale factor that creates a non-singular bounce but rather comes from the interplay between the modified gravity and the matter field dynamics. Therefore, the scale factor that we have chosen acts as a scientifically grounded background that enables us to systematically explore the effects of our theoretical framework on the bouncing evolution.}
\section{Model-I: Bouncing behaviour in the phantom scalar field with Gauss-Bonnet term}\label{2}
\begin{figure}[!ht]
    \centering
    \includegraphics[width=0.5\linewidth]{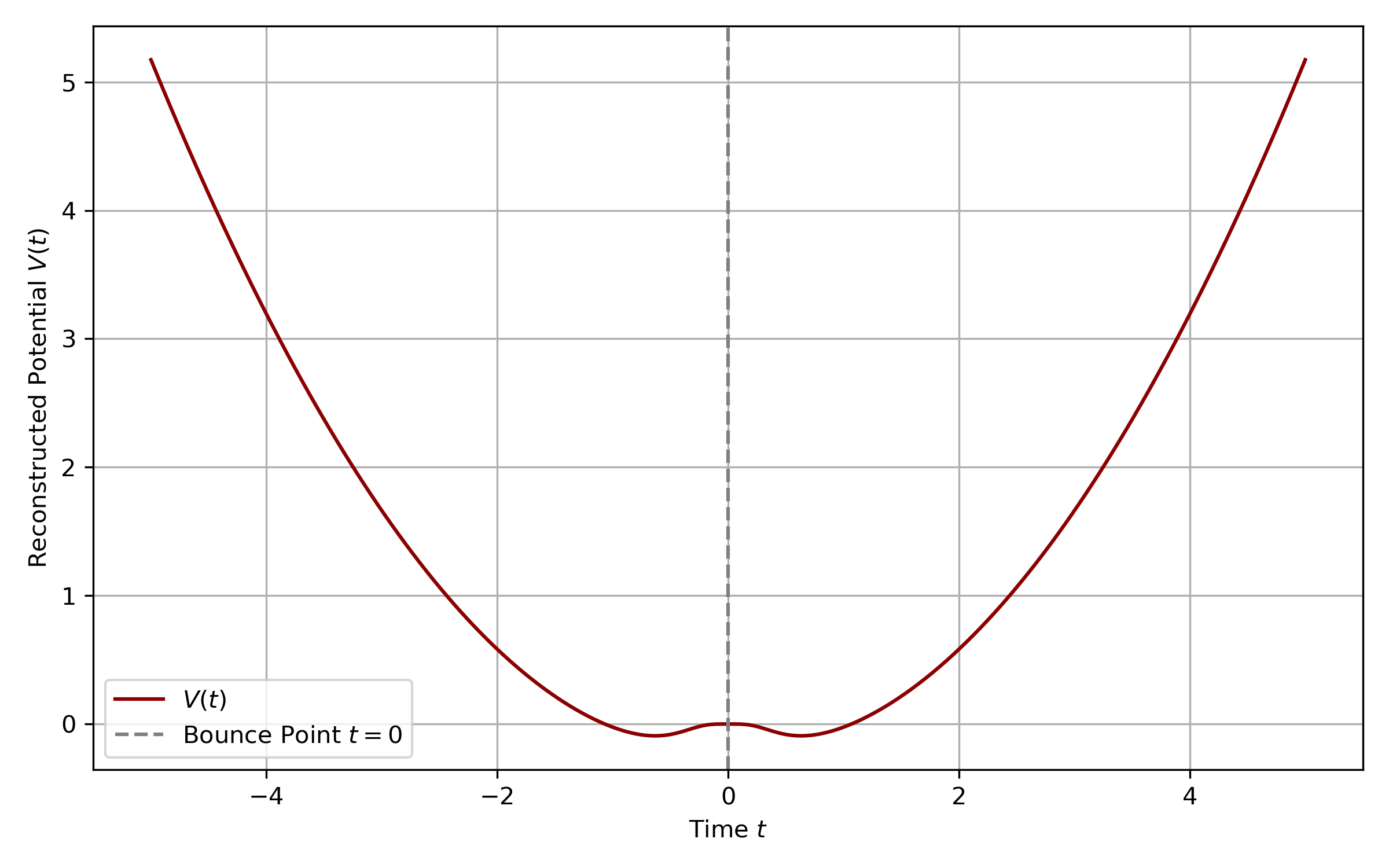}
        \caption{\textcolor{black}{Behaviour of the reconstructed $V(t)$ plotted against cosmic time $t$. We have taken $\phi_{0} = 0.07$,  $n= 2$, $f_{0}= 0.04$, $h= 0.06$, $\rho_{m_0}= 0.45$, $\alpha=0.2506, \eta=1.0924$, $m= 0.08$, $C_{1}= 10.6$, $\lambda=0.008$, $B=1.006$ and $\rho_{c}= 0.05$.}}
    \label{F1a}
\end{figure}

The observations of the bounce-inspired scale factor discussed in the previous section provides us a strong motivation for its use in reconstructing the scalar potential $V(t)$. The analytic compliance of the scale factor guarantees closed-form expressions for the Hubble parameter and its derivatives, which are essential for solving the reconstruction equations.
Thus, we proceed with this scale factor to analyze the bouncing scenario while considering the phantom scalar field with the Gauss-Bonnet term.  Eqns. (\ref{F2},\ref{F3},\ref{F4}) give the field equations for this field. In order to solve for $V(\phi)$, we have assumed that $\phi=\phi_0t^n$ and $f(\phi)=f_{0}\phi=f_{0}\phi_{0}t^{n}$ which is the minimal non-trivial choice that allows the Gauss--Bonnet term to influence the dynamics without introducing higher-derivative instabilities. Here $n$ is a positive constant. Thus, $f_{,\phi}=\frac{df}{d\phi}=f_{0}$ and $f_{,,\phi}=0$. Additionally, we have considered $g(\phi)=Be^{\lambda\phi_{0}t^{n}}$ which can be expanded to give 
\begin{equation}
   g(\phi)=B\left(1+\lambda\phi_{0}t^{n}+\frac{\lambda^2\phi_{0}^2t^{2n}}{2!}+\frac{\lambda^3\phi_{0}^3t^{3n}}{3!}+...\right)
\end{equation}
which expresses the coupling explicitly in terms of cosmic time. Near the bounce, $t$ is small, so higher-order terms in the exponential contribute very little. This allows us to safely truncate the series as
\begin{center}\label{r1}
   $g(\phi)=B\left(1+\lambda\phi_{0}t^{n}+\frac{\lambda^2\phi_{0}^2t^{2n}}{2!}\right)$ 
\end{center} 
 Unlike \cite{papagiannopoulos2025avoiding}, where an exponential potential 
\begin{equation}
V(\phi) = V_0~ e^{\lambda \phi}
\end{equation}
was assumed, we instead take an ansatz for the scalar field,
\begin{equation}
\phi(t) = \phi_0 \, t^n,
\end{equation}
and reconstruct the potential $V(t)$ directly from the field equations. This reversed strategy is equally consistent: it allows us to derive a closed-form analytic potential well governing the bounce, while still retaining the link 
\begin{equation}
V(\phi) = V_0 ~ \phi
\end{equation}
through the chosen ansatz. These choices are therefore not arbitrary but motivated by earlier work \textcolor{black}{\cite{papagiannopoulos2025avoiding}}, analytic tractability, and the requirement of a regular bounce, while our adaptation extends the framework to suit the reconstruction approach adopted here. \textcolor{black}{Furthermore, the change of the scalar field around the bounce is controlled by the exponent $n$. Contraining of it is needed for regularity and stability. Since the bounce happens at $t=0$, picking $n>0$ keeps the scalar field from blowing up at the bouncing point and avoids the triviality of constant field configurations. Additionally, in order to have both the first and second time derivatives of $\phi(t)$ staying finite close to the bounce, we need $n$ to impose $n\geq2$. This way, the scalar field evolves smoothly, and we do not run into any problems in the Gauss–Bonnet coupling terms.
With $n$ in this range, the Taylor expansion for the exponential coupling $g(\phi)$ in \eqref{r1} actually stays well-defined and converges around the bounce. Thus, truncating off the series where we do makes sense. In short, we treat $n$ as a positive constant, big enough for regularity, and not just some phenomenological parameter for a better fit.} \textcolor{black}{However, we note that, in contrast to the papers found in the literature that employ such reconstruction techniques \cite{paliathanasis20244d, papagiannopoulos2025avoiding} wherein the potential is presumed a priori, our method reconstructs $V(t)$ straight from the field equations, considering the phantom scalar field dynamics and Gauss–Bonnet coupling. Thus, our reconstruction strategy leads to the consideration of diverse impacts induced by the interaction between the scalar field, Gauss–Bonnet term, and matter coupling, such as the small potential asymmetry around the bounce and alterations due to viscous and non-viscous matter components. Therefore, our method offers a new analytic control over the bounce evolution and related cosmological observables.}
Using the ansatzes as mentioned earlier, the effective fluid now becomes :
\textcolor{black}{\begin{equation}\label{b3}
 \rho_{eff}=-\frac{1}{2}Q^2~t^{2(n-1)}+V(\phi)+B\left(1+St^{n}+\frac{S^2t^{2n}}{2!}\right)\rho_{m_{0}}~a^{-3}+24 R H^2 n t^{n-1}
\end{equation}}
and
\begin{equation}\label{b4}
   \textcolor{black}{p_{eff}=-\frac{1}{2}Q^2~t^{2(n-1)}-V(\phi-16 (H^2+\dot{H})HRnt^{n-1}-8H(Rn(n-1)t^{n-2})}
\end{equation}
\textcolor{black}{From here onwards, we will take $Q=\phi_0~n$, $R=f_0 \phi_0$ and $S=\lambda \phi_0$} in our work.
Furthermore, Eq. (\ref{F4}) becomes:
\begin{equation}\label{b5}
    0=\frac{\dot{V}}{Q t^{n-1}}-3HQ t^{n-1}-Pt^{n-2}+24R(H^2+\dot{H})+B\left(1+\lambda\phi_{0}t^{n}+\frac{\lambda^2\phi_{0}^2t^{2n}}{2!}\right)\rho_{m_{0}}~a^{-3}
\end{equation}
 Eq. (\ref{b5}) was then used to solve for $V(t)$:
 \textcolor{black}{
\begin{equation}\label{V1}
\begin{aligned}
V(t)&= C_1 
+ \frac{Q\, t^{\,n-2}\! \left(t^2 + \frac{\alpha}{\eta}\right)^{-\tfrac{3}{2\eta}}}
{2(1+n)(2+n)\,\alpha^3 \eta}\\
&\Bigg[
-3n^2(2+n)Q\,t^{4+n}\alpha
\left(t^2+\frac{\alpha}{\eta}\right)^{\tfrac{3}{2\eta}}\eta^2
\,{}_2F_1\!\left(1,1+n;2+n;-\tfrac{t^2\eta}{\alpha}\right)
\\[6pt]
&+(1+n)\Big(
48R\,n\,t^4\!\left(t^2+\tfrac{\alpha}{\eta}\right)^{\tfrac{3}{2\eta}}(-1+\eta)
\,{}_2F_1\!\left(3,1+\tfrac{n}{2};2+\tfrac{n}{2};-\tfrac{t^2\eta}{\alpha}\right)
\\[4pt]
&-48R\,n\,t^4\!\left(t^2+\tfrac{\alpha}{\eta}\right)^{\tfrac{3}{2\eta}}(-1+2\eta)
\,{}_2F_1\!\left(4,1+\tfrac{n}{2};2+\tfrac{n}{2};-\tfrac{t^2\eta}{\alpha}\right)
\\[4pt]
&+(2+n)\alpha^2\eta\Big(
-2B\rho_m t^2\alpha\!\left(1+\tfrac{t^2\eta}{\alpha}\right)^{\tfrac{3}{2\eta}} 
S\,{}_2F_1\!\left(\tfrac{n}{2},\tfrac{3}{2\eta};1+\tfrac{n}{2};-\tfrac{t^2\eta}{\alpha}\right)+ Q\,t^n
\\[4pt]
&\!\Big[n(3t^2+n\alpha)\!\left(t^2+\tfrac{\alpha}{\eta}\right)^{\tfrac{3}{2\eta}}
- B\rho_m t^2\alpha\!\left(1+\tfrac{t^2\eta}{\alpha}\right)^{\tfrac{3}{2\eta}}
S^2\,{}_2F_1\!\left(n,\tfrac{3}{2\eta};1+n;-\tfrac{t^2\eta}{\alpha}\right)
\Big]\!\Big)\!\Big)\!\Bigg]
\end{aligned}
\end{equation}
}
 Fig. \ref{F1a} shows this reconstructed potential plotted against $t$,  using the best-fit parameters \textcolor{black}{$\alpha = 0.2506$ and $\eta = 1.0924$} from the observational analysis(see Section\ref{6}). It may be noted that the scalar field dynamics driving the bounce is expressed by the potential well that is observed in the figure which shows a \textcolor{black}{slight asymmetry around the bouncing point $t = 0$.} The dashed vertical line signifies the moment at which the transitions from a contracting phase to an expanding phase occurs. \textcolor{black}{The minimum of $V(t)$ is not reached at $t = 0$. Instead, we note that the minimum is slightly shifted, which tells us that the scalar-field dynamics undergoes some sort of imbalance during the two phases (contracting and expanding). However, the} non-singular behaviour of the bounce and the lack of instabilities in the scalar framework are suggested by the curve's regularity and smoothness. Furthermore, the reconstructed form of $V(t)$ contains both the geometric and thermodynamic aspects of the bounce while including hypergeometric contributions from the Gauss-Bonnet term and the scalar field coupling. Thus, in our work, the slow-roll analysis and predictions in the $n_s$–$r$ plane, which we have described in the subsection \ref{sub1}, are based on this potential, in an effort to connect theory to observations.

\begin{figure}[!ht]
    \centering
    \begin{subfigure}[b]{0.45\textwidth}
        \centering
\includegraphics[width=1.1\linewidth]{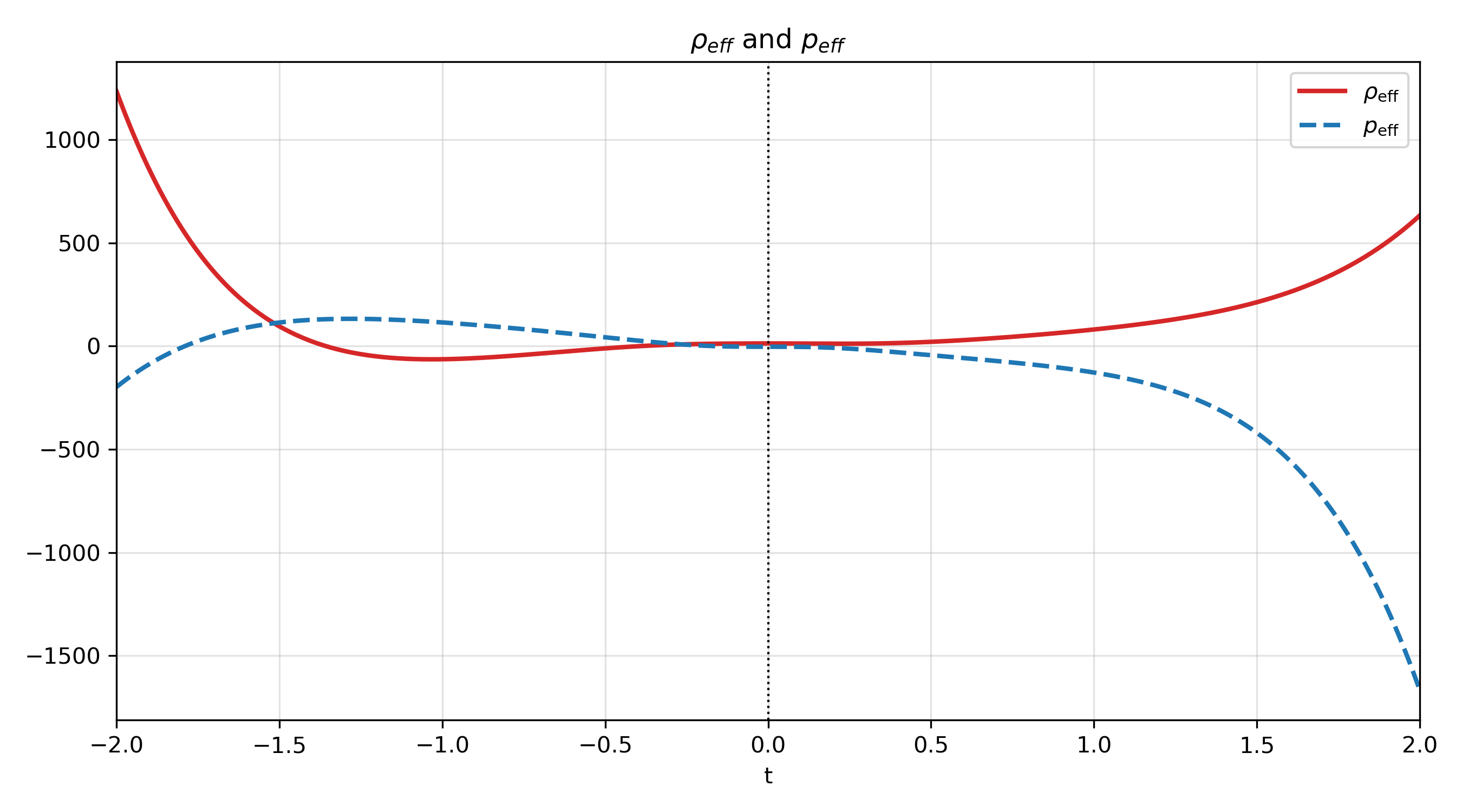}
        \caption{\textcolor{black}{The reconstructed density and pressure plotted against cosmic time. Here, we have taken $\phi_{0} =1.01$, $n=4$, $f_{0}=1.5$, $\rho_{m_0} =0.8$, $\alpha=0.2506, \eta=1.0924$, $B=1.60$, $\lambda=0.91$ and $C_{1}= 0.91$}}
        \label{rho}
    \end{subfigure}
\hfill
    \begin{subfigure}[b]{0.35
\textwidth}
        \centering
\includegraphics[width=1.1\linewidth]{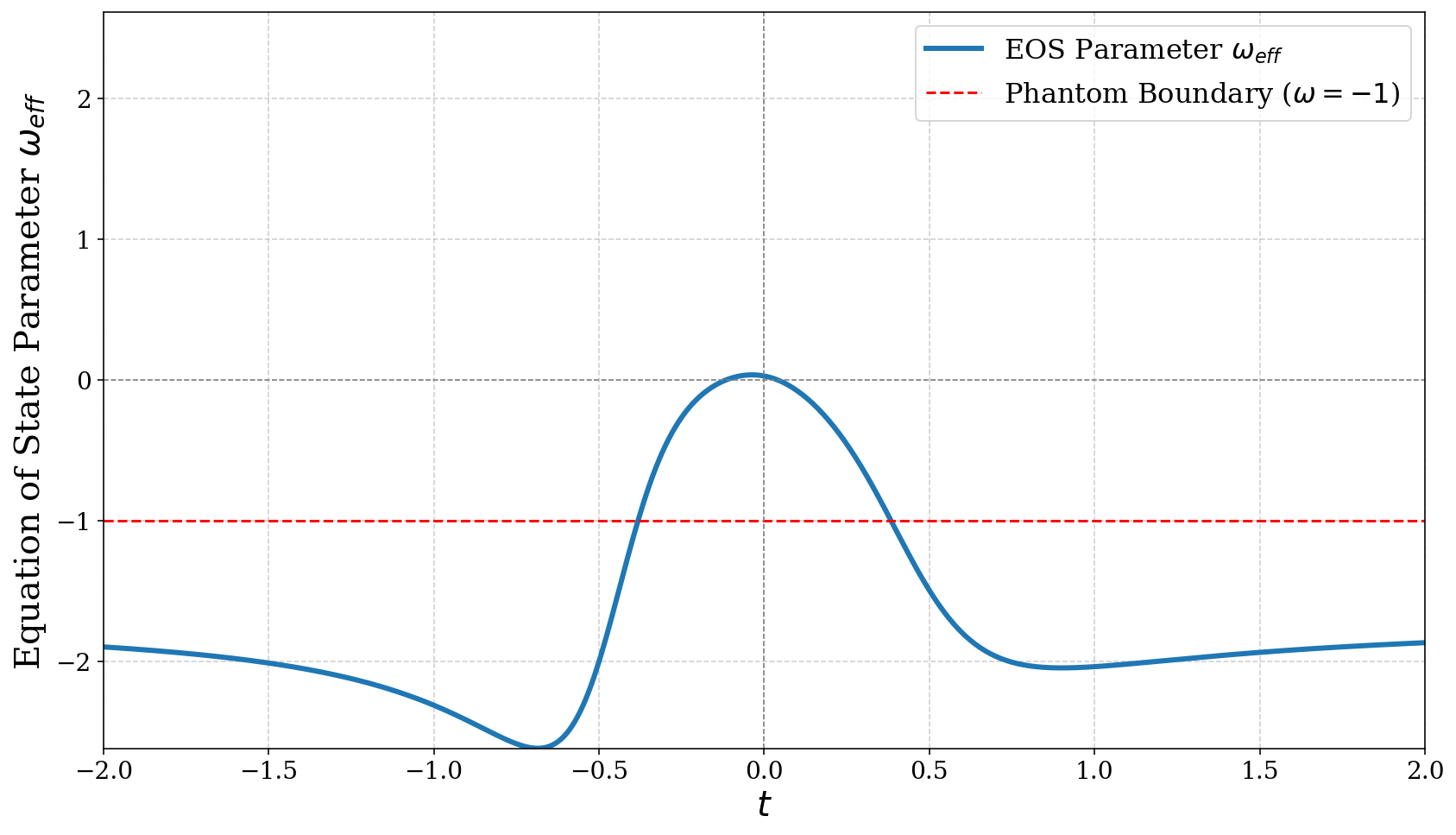}
        \caption{\textcolor{black}{Behaviour of the reconstructed EoS parameter $\omega$ plotted against cosmic time $t$ where we have taken $\phi_{0} =0.5$, $n=2$, $f_{0}=0.005$, $\rho_{m_0} =0.5$, $\alpha=0.2506, \eta=1.0924$, $B=0.11$, $\lambda=0.1$ and $C_{1}=-0.01$}}
        \label{eos}
    \end{subfigure}
    \caption{Behaviour of the reconstructed density and  EoS parameter($w$) plotted against cosmic time $t$ for model-I.}\label{F1b}
    \end{figure}

The reconstructed $V(\phi)$ function was substituted in Eq. \ref{b3} to obtain the reconstructed density parameter. Similar substitution was made in Eq. (\ref{b4}) which leads to the reconstructed pressure. Finally, the reconstructed EoS parameter containing the terms of the field and the bouncing scale factor was obtained. The reconstructed density and pressure plotted against cosmic time $t$ are displayed in subfig. (\ref{rho}) of Fig. \ref{F1b}. We have used the best-fit parameters $\alpha=0.2506$ and $\eta=1.0924$. The energy density $\rho(t)$ curve \textcolor{black}{shows an asymmetric behaviour. It begins at high values in the contracting phase and becomes slightly negative at around $-1.2\leq t\geq-0.3$. As seen in the figure, there's a shallow minimum that is reached at around $t=-1$ after which it rises again and crosses 0 at the bouncing point $t=0$. Following this, it continues to increase in the expanding ($t > 0$) phase. This transient negative behaviour of density is expected within the context of effective fluid reconstruction in Gauss-Bonnet gravity models, specifically in the case of bounce, as a violation of NEC is essential for a smooth and non-singular bounce. Furthermore, the pressure $p(t)$ (blue dashed line) seems to show complementary behaviour to the density parameter. It is asymmetric and starts at negative values. It continues to increase and reaches 0 near $t\approx-0.5$ and crosses $t=0$ at the bouncing point. The pressure curve also reaches a small positive maximum between $-1.5\leq t\geq -0.5$. In the post-bounce($t>0$), we see that the pressure becomes rapidly negative, which is an indication of the strong repulsive effect brought on by the scalar Gauss-Bonnet coupling in the early expansion era. The simultaneous crossing of $t=0$ exhibited by both $\rho_{eff}(t)$ and $p_{eff}(t)$ highlights the transition from contraction to the expansion phase and the non-singular nature of the bounce.}\\

The subfigure~\ref{eos} of Fig.~\ref{F1b} shows the behaviour of the reconstructed equation of state (EoS) parameter $\omega(t)$ as a function of cosmic time. Evidently, the curve is \textcolor{black}{asymmetric and shows oscillatory behaviour across the bouncing regime.}It is important to mention here that for our model, the evolution of $\omega(t)$ spans both the phantom ($\omega < -1$) and quintessence ($\omega > -1$) regions. \textcolor{black}{During the pre-bounce phase, $\omega(t)$ is below $-1$ and remains in the phantom region, with a minimum value of approx. $2.5$ near $t\approx-0.75$. Thereafter, the EoS parameter continues to increase monotonically and crosses the phantom divide $\omega=-1$ around $t\approx-0.5$ and enters the quintessence regime. Additionally, it crosses the pressureless state slightly, i.e. $\omega=0$ just before the bounce. Another point of significance is that the curve attains the value $\omega=0$ at the bouncing point, which signifies an ephemeral pressureless state during the transition from the contracting to the expanding phase. However, in the post-bounce era, we observe that the EoS immediately dips below  $\omega(t)=0$ and crosses the phantom divide again at  $t\approx0.5$, transitioning smoothly into the phantom region and remains in this region. Thus, the reconstructed effective fluid exhibits two phantom divide crossings and this oscillatory behaviour tells us that the effective cosmic fluid behaves as a dynamic dark energy component that is not constant and is driven by the dynamics between the scalar-field, coupling and the Gauss-Bonnet term.} Thu, we can say that the absence of either discontinuities or singularities strengthens the physical viability of the model.

\section{Model-II: Bouncing behaviour in phantom-scalar field model with Gauss-Bonnet term while considering bulk viscosity}\label{3}
The limitations of perfect fluids, particularly when perturbations are present, have served as a motivator for exploring an imperfect equation of state in a variety of cosmological models. This enables treatment of the Eos parameter, and hence the model, in a more realistic manner. In such a cosmological scenario, two coupled components: dark matter with a linear homogeneous EOS and dark energy with a linear inhomogeneous EOS are considered. \cite{brevik2014bounce} is an example of a study of cosmological bounce in a viscous setting. Other studies on the interaction between dark energy and dark matter in inhomogeneous fluids can be found in \cite{brevik2013little,timoshkin2009specially}.\\
Among myriad formulations of imperfect fluids, Van Der Waal's (VDW) fluid \cite{Brevik2018, Brevik1999, Brevik2003, Sadhukhan2021, Obukhov2018} is a compelling example, and in cosmology, the VDW equation of state is formulated on the assumption that dark matter and dark energy are one fluid with non-ideal behaviour. Models of VDW fluid can account for both inflation and late-time acceleration. The studies related to VDW cosmology can be found in \cite{kremer2003cosmological,capozziello2003quintessence,kremer2004brane,khurshudyan2014interacting}.
In this section, we discuss the viscous (i.e. bulk viscosity contribution is taken into consideration) bouncing model wherein we consider the bouncing scale factor we have dealt with so far and VDW fluid with the equation of state \cite{Brevik2017}:
\begin{equation}\label{v1}
  p_{eff}=w(r,t)\rho+f(\rho)-3 H \xi(H,t)  
\end{equation}
where $w(r,t)$ is a thermodynamic parameter depending on $r=\frac{\rho_{m}}{\rho}$ and $t$. Additionally, $\xi(H,t)$ is the bulk viscosity term that is dependent on the Hubble parameter and time $t$. Following \cite{Brevik2017}, we have assumed that the VDW has a parametrized EoS and the following forms of $f(\rho)$ and the viscosity:
\begin{equation}\label{v2}
    w(r,t)= \frac{\gamma}{1-\beta \left(\frac{\rho}{\rho_c}\right)}= \frac{\gamma}{1-\beta \left(\frac{\rho_m}{r\rho_c}\right)}~~~~~~\left[\because r=\frac{\rho_m}{\rho}\right]
\end{equation}

\begin{equation}\label{v3}
   f(\rho)=-\frac{\sigma}{\rho_c}\rho^2
\end{equation}
\begin{equation}\label{v4}
  \xi(H,t)=\tau(3H)^m
\end{equation}
In the above equations, $\gamma$, $\beta$ and $\sigma$ are independent parameters, whereas $\rho_c$ is the critical density at which the phase transition occurs and $\rho_m=\rho_{m_0}a^{-3}$ Additionally, $\tau$ is a positive constant for VDW fluid and $m$ is the power dependency of the viscosity on $H$. We have chosen $m=2$. Let us now justify the use of $m=2$. The exponent \( m \) in Eq.~\eqref{v4} indicates the strength and evolution of the bulk viscosity's contribution to the cosmic dynamics. Choosing \( m = 2 \) provides a physically motivated and mathematically tractable form that ensures the viscosity grows quadratically with the Hubble parameter. As we are taking $m=2$, there will be a highly significant bulk viscosity near the bounce and it would suppress the anisotropy to make it a non-singular bounce. Moreover, \( m=2 \) is consistent with previous studies that successfully modeled bounce and late-time acceleration within modified gravity and viscous fluid frameworks (see e.g.,\cite{Brevik2017}). This quadratic form also allows a smooth decay of viscosity with the expansion of the universe and is expected to be consistent with the post-bounce cosmology.

At this juncture, let us have some insight into the expressions shown above. The inclusion of bulk viscosity in cosmological fluid dynamics introduces essential dissipative effects, particularly relevant near bounce epochs. The generalized equation of state in Eq.~\eqref{v1} has a time- and ratio-dependent barotropic term $w(r,t)$, a nonlinear correction $f(\rho)$, and a viscous pressure term $-3H\xi(H,t)$. Within the VDW framework, parameters $\gamma$ and $\beta$ imply non-ideal behavior and critical phenomena. As seen in Eq.~\eqref{v2}, the dependence of $w(r,t)$ on the ratio $r = \rho_m / \rho$ introduces a dynamic matter-energy coupling crucial to bounce dynamics. The viscosity function $\xi(H,t)$ generates negative pressure, enabling singularity avoidance and stabilizing the bounce \cite{Brevik2017}. Coupled with a bouncing scale factor, this viscous VDW model offers a realistic and thermodynamically consistent fluid, helping in a smooth transition into the expanding phase. 

\begin{figure}[!ht]
    \centering
    \begin{subfigure}[b]{0.35\textwidth}
        \centering
\includegraphics[width=1.1\linewidth]{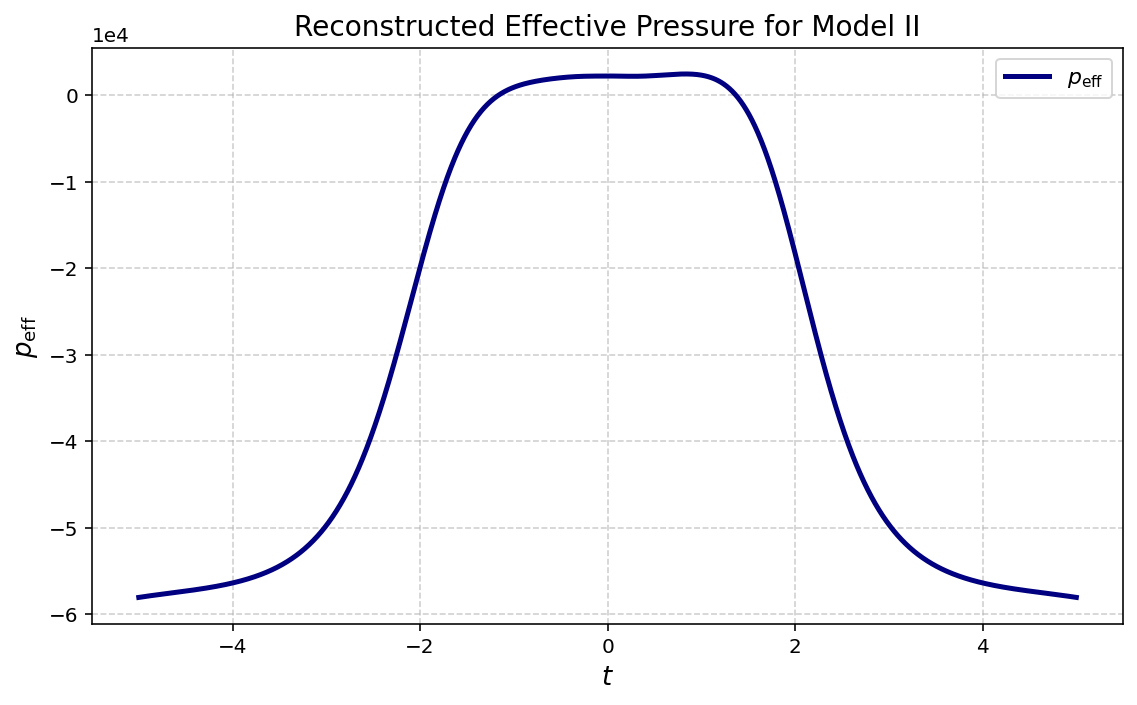}
        \caption{\textcolor{black}{The reconstructed pressure plotted against cosmic time $t$ while considering bulk viscosity.}}
        \label{rhpvo}
    \end{subfigure}
\hfill
    \begin{subfigure}[b]{0.35
\textwidth}
        \centering
\includegraphics[width=1.1\linewidth]{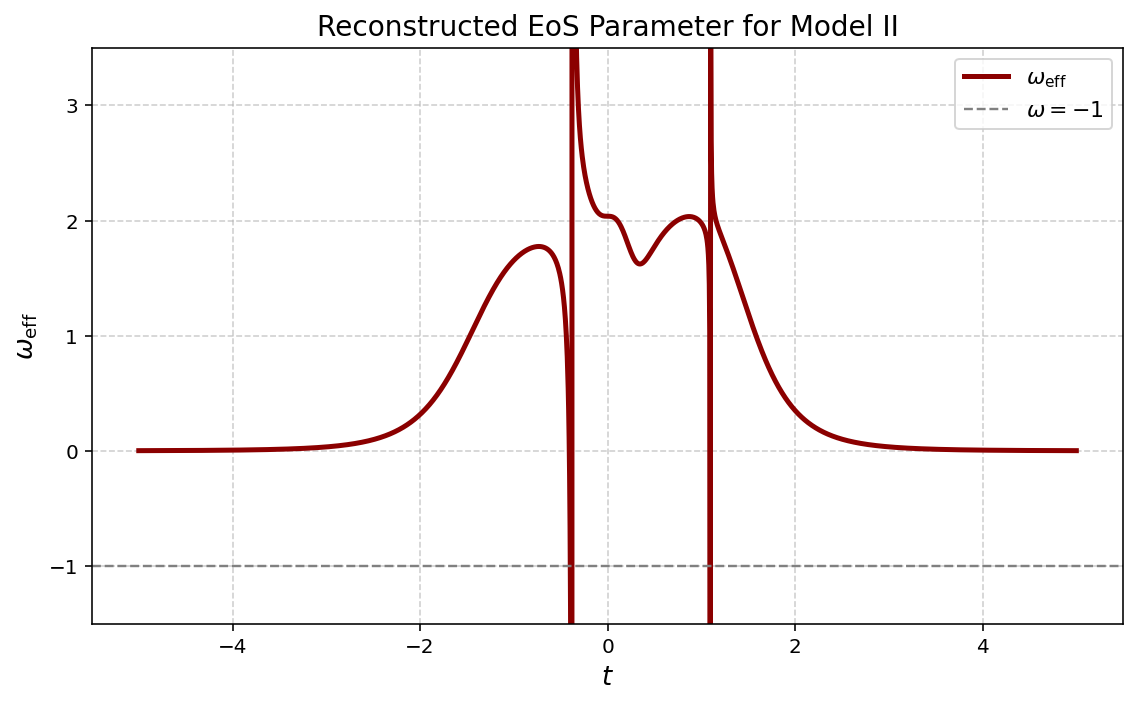}
        \caption{\textcolor{black}{Behaviour of the reconstructed EoS parameter $\omega$ plotted against cosmic time $t$ for model II that incorporates a bulk viscosity term $\xi \propto H^2$. The two thin, vertical spikes indicate the mathematical singularities where the effective energy density $\rho_{\mathrm{eff}}$ vanishes.}}
        \label{eosv}
    \end{subfigure}
    \caption{The reconstructed pressure and the EoS parameter($\omega$) plotted against cosmic time $t$ in case of a bouncing model while considering bulk viscosity. In this case, our chosen values of the parameters are $\phi_{0}=0.107, m=2, f_{0}=1.04, h=0.06, \rho_{m}=0.45, \alpha= 0.2506, \eta= 1.0924, m=1.08, C_{1}=1.92, \rho_{c}=2.95, \beta=2.89, \gamma=1.0057, \tau=0.005, \xi=3.004, \sigma=3.067$}\label{F2a}
    \end{figure}

Employing the reconstructed potential $V(t)$ given in Eq. \eqref{V1} and our reconstructed effective density in the phantom scalar field with the Gauss-Bonnet term described in the previous section,
Eq. (\ref{v1}) was used to derive the reconstructed pressure for the viscous model. \textcolor{black}{Fig. \ref{F2a} (left panel) shows the reconstructed effective pressure plotted against the cosmic time $t \in [-2, 2]$ while constraining the $\alpha$ and $\eta$ values within the observational bounds. We see that the curve is a highly symmetric bell-shaped curve that is inverted and centred around the bouncing point $t=0$. The pressure $p(t)$ remains largely negative and plateaus in the neighbourhood of the bouncing point, reaching a positive value slightly greater than 0, and then immediately starts decreasing again after it crosses $t=0$. In comparison to the non-viscous model, this model seems to show more symmetry and never becomes significantly positive, indicating that the bulk viscosity regulates the pressure evolution, suppressing instabilities, oscillatory features and reinforcing the bounce’s regularity. In summation, the reconstructed pressure for the viscous model shows a well-regulated and robust bouncing scenario, consistent with a dark-energy-like fluid/component.\\
Then, we have computed the equation of state parameter ($\omega=\frac{p_{eff}}{\ rho_{eff}}$) and these values were plotted against cosmic time $t$ as shown in Fig. (\ref{F2a}) (right panel). As we can see, the evolution is distinctly different from the case of the non-viscous model. The significant difference is that for model-II, there are violent variations near the bounce. For $|t|\geq0$, the EoS parameter is flat and reaches 0 asymptotically, which is an indication that the fluid will behave like pressureless matter away from the bounce. When we focus on the central region ($-3\leq t\geq3$), we observe that the $\omega_{eff}(t)$ significantly departs from the pressureless state. In the contracting phase, the curve rises from the asymptotic region($\omega_{eff}(t)=0$) and reaches a maximum of around 1.8 around $t\approx -0.75$. Immediately after($t\approx -0.3$), the EoS parameter shows a vertical singularity and diverges to $\pm\infty$, which can be attributed to the zero crossing of the effective energy density which renders the ratio $\frac{p_{eff}}{\rho_{eff}}$ undefined. Following this, the behaviour of $\omega_{eff}$ shows rapid and finite. oscillations, adopting values between 2.2 and 1.7. A second singularity is seen around $t\approx 1.25$, after which the value rapidly declines and assumes a dust-like behaviour. Thus, we note that throughout the evolution, except for the singularities, the EoS parameter remains in the quintessence region, which is in stark contrast to the non-viscous model discussed in the previous section. The divergences arise purely from the vanishing of $\rho_{eff}$. Thus, we can conclude that the bulk viscosity largely alters the effective dynamics while avoiding phantom behaviour and confining it to a quintessence-like behaviour with brief periods of null-density singularities.}

  \section{Energy Conditions}\label{4}
\begin{figure}[!ht]
    \centering
    \begin{subfigure}[b]{0.48\textwidth}
        \centering
\includegraphics[width=\textwidth]{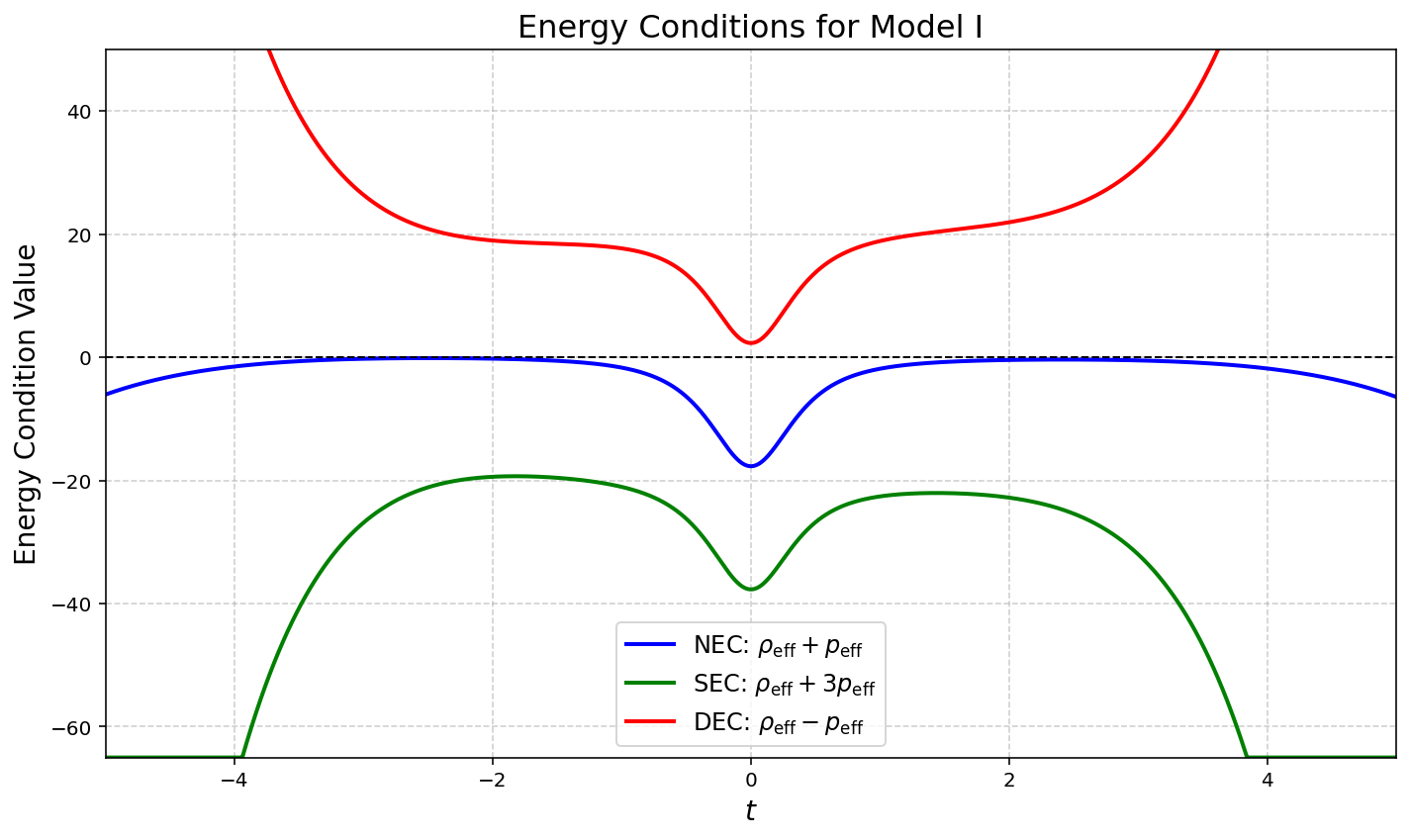}
        \caption{\textcolor{black}{Evolution of energy conditions versus cosmic time for the bounce model without bulk viscosity. In this case, we have taken $B=1.78, \phi_{0}=0.07, n=2, f_{0}=2.24, \rho_{m}=0.45, \alpha=0.2506, \eta=1.0924, C_{1}=0.04$}
}
        \label{n1}
    \end{subfigure}
\hfill
    \begin{subfigure}[b]{0.48
\textwidth}
        \centering
\includegraphics[width=\textwidth]{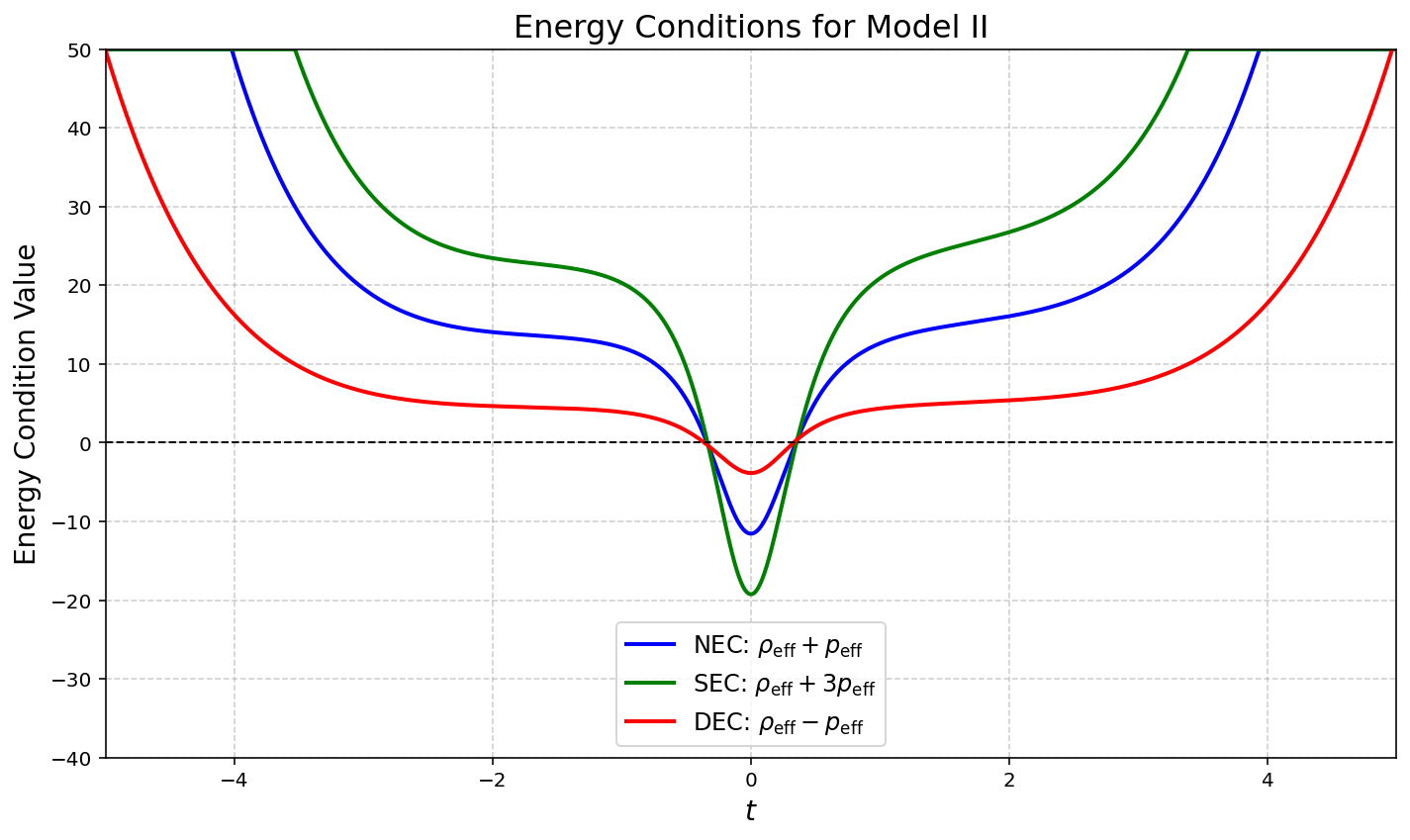}
        \caption{\textcolor{black}{Evolution of energy conditions versus time for the bulk-viscous bounce model where the parameter were taken as $B=1.1160, \phi_{0}=0.005, m=4, f_{0}=0.955, \rho_{m0}=2.101, \alpha=0.2506, \eta=1.0924, C_{1}=1.01, \rho_{c}=2.9, \beta=0.0001, \gamma=0.50, \tau=0.01, \xi=1.04, \sigma=1.067$
}}\label{s1}
    \end{subfigure}
    \caption{Energy conditions plotted against cosmic time $t$ while considering the bouncing models with and without bulk viscosity.}
    \label{F3a}
    \end{figure}

Significant insights into the cosmological geometries and gravitational fields can be brought to light by exploring the energy conditions of general relativity(GR). Einstein's theory of gravity which relates geometry to matter, is undoubtedly a complex theory, even under the restriction of just the classical regime. The left-hand side of the field equation:
\begin{equation}
  G^{\mu\nu}=\frac{8\pi G_{Newton}}{c^4}T^{\mu\nu} 
\end{equation}
which is the Einstein tensor and describes the geometry is comparatively less complex due to the universality of spacetime geometry. However, the stress-energy tensor, $T^{\mu \nu}$ is dependent on the matter type one considers and the interactions assumed in the model which leads to numerous possible cases, one model per matter Lagrangian that can be written down. Alternately, generic attributes can be developed that must be satisfied by all apt stress-energy tensors. These can then be utilised to produce universal theorems that can be applied to describe the behaviour of strong gravitational fields. The (almost) invariably positive density is an example of one such generic feature and physicists use what is called "energy conditions" which impose mathematical requirements on certain combinations of $T^{\mu\nu}$ that make the notion of positivity of energy density more precise. These conditions are pointwise conditions i.e. they should hold at every point and not just on average. 
These point-wise energy conditions are as follows:
\begin{equation}
Null~Energy~Condition(NEC):~\rho + p \geq 0
\end{equation}

\begin{equation}
Weak~Energy~Condition(WEC):~ \rho \geq 0,~~~~~~ \rho + p \geq 0
\end{equation}

\begin{equation}
Strong~Energy~Condition(SEC):~\rho + 3p \geq 0
\end{equation}

\begin{equation}
Dominant~Energy~Condition(DEC):~\rho \geq |p|
\end{equation}
The NEC is the weakest and it is directly associated with the imposition on the equation of state parameter $\omega$. Experimental data allows for the violation of NEC ($\omega<-1$) and such models have attracted a lot of interest. The NEC violation is a prerequisite for bounce to happen. Furthermore, it has been noted that all other pointwise energy conditions would likewise be broken if the null energy condition (NEC) was broken. However, even if the NEC is broken, the dominant energy condition (DEC) is satisfied for dark energy dominated models with a cosmic fluid that has a negative pressure. Thus, as in our study, an analysis of the energy conditions for any bouncing model is required.\\ Specifically, in our study, the bouncing scale factor given by Eq. (\ref{b1}), is employed inside the phantom scalar field models we have explored so far. This is done to reconstruct the potential of the scalar field, energy density and pressure for each model to plot the graphs of the various energy conditions against cosmic time as shown in Fig. \ref{F3a}
As depicted in the left panel of Fig. \ref{F3a}, for model-I, i.e. the phantom scalar field model with Gauss-Bonnet gravity, without considering bulk viscosity, the NEC \textcolor{black}{stays around 0 over most of the domain but dips below it between $t\approx -1.5$ and $t\approx 1.5$, reaching a minimum at the bouncing point $t=0$. This violation of NEC is what enables a smooth, non-singular bounce. One may note that the validation of the SEC implies attractive gravitational forces. The SEC, on the other hand, is the most negative curve and is violated throughout, which tells us that the pressure is sufficiently negative to generate repulsive gravity away from the bounce. On the contrary, the DEC is satisfied throughout and forms a broad U-shaped profile with a shallow minimum at $t=0$, which means that the fluid respects causal behaviour. Taken together, the overall pattern demonstrates that the model meets the preliminary causal requirements, the minimal conditions for a smooth bounce and maintains a repulsive gravitational scenario that is essential for late-time acceleration.}\\

We observe that when bulk viscosity is included in the second model, the evolution of the Null Energy Condition (NEC: $\rho + p$), Strong Energy Condition (SEC: $\rho + 3p$), and Dominant Energy Condition (DEC: $\rho - p$) across cosmic time display a different behaviour than the viscous model as shown in subfig. \ref{s1} of Fig. \ref{F3a}. \textcolor{black}{The DEC curve remains positive and satisfied throughout the time domain, except when it dips below zero for a tiny region around the bounce. This signifies a transient violation where the energy flow becomes non-causal. Another notable difference in the energy conditions of the two models is that the NEC and SEC curves stay positive through most of the time domain and reach negative values around the bouncing epochs. They attain their minimum values at  $t=0$ and then rise again to become positive in the post-bounce epoch. These violations are small and brief and are resolved quickly, which tells us that the inclusion of bulk viscosity affects the fluid dynamics and ensures that the bounce is realised without large departures from the standard energy conditions.} This difference between the NEC and SEC violation and the extended DEC satisfaction proves that the viscous model stays physically acceptable yet creates unique post-bounce dynamics in contrast with the non-viscous scenario.\\
In conjunction, the energy condition profiles of both models point to a very delicate balance needed for a physically viable bounce. The contrast in transient-sustained violations underlines the possibility for bounce cosmologies to realise various mechanisms--be they geometric or dissipative--while conforming to the main energy constraints. These results suggest a likely positive role of dissipative processes like bulk viscosity in early-universe dynamics, leading to still more robust and observationally consistent bouncing scenarios.

\section{Stability Analysis}\label{5}
\begin{figure*}[!ht]
        \centering
\includegraphics[width=0.8\linewidth]{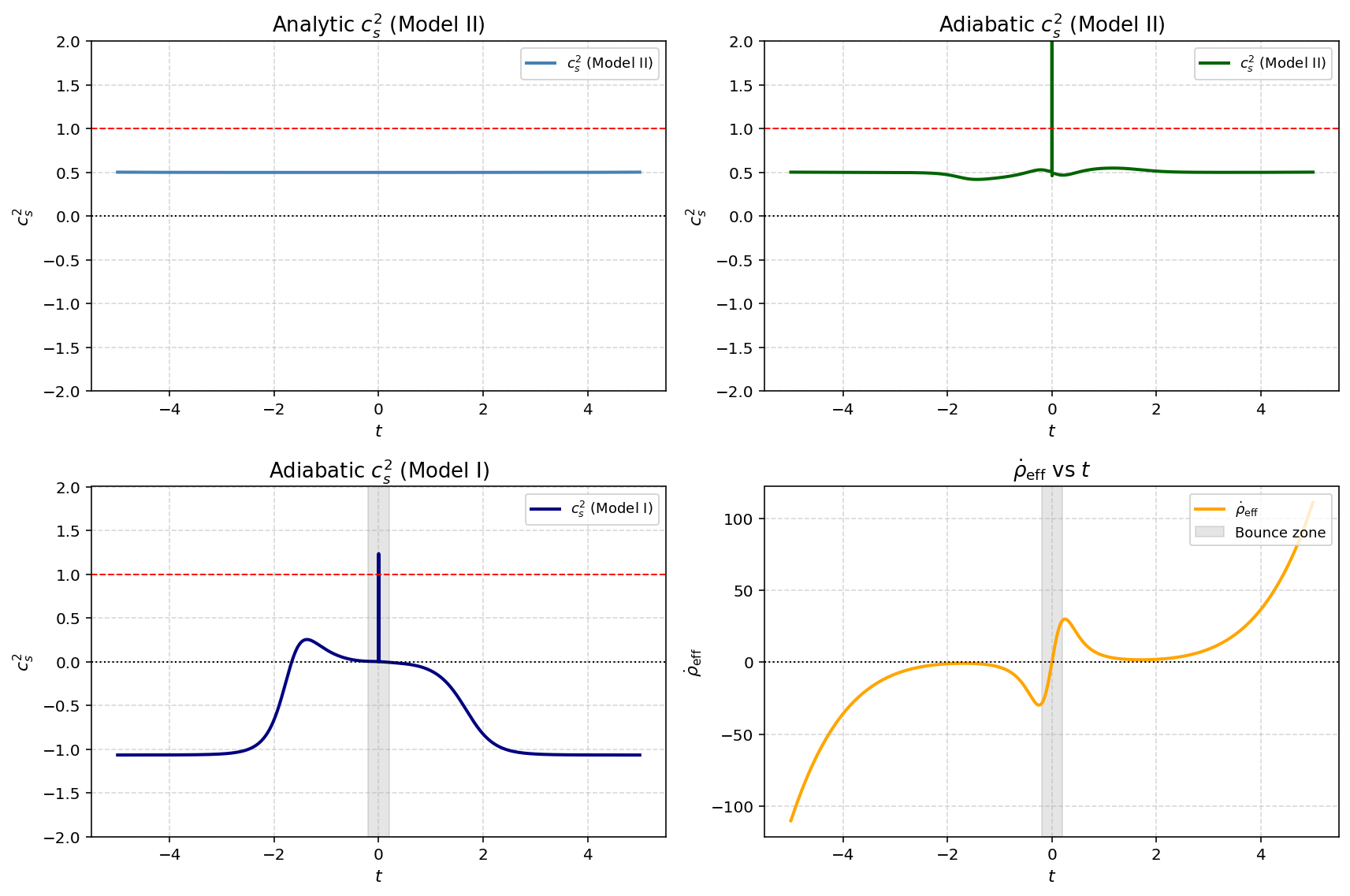}
    
    \caption{Variation of the squared speed of sound for the non-viscous and viscous models, to assess classical stability. Here, \textcolor{black}{$\phi_{0}=0.005, n=4, f_{0}=0.955, B=1.1160, \rho_{m_0}=2.101, \alpha=0.2506, \eta=1.0924, C_{1}=1.01, \tau=0.01, \rho_{c}=2.9$ and $\sigma=1.67$}}
    \label{F5a}
    \end{figure*}

Apart from the equation of state parameter, the speed of sound is also a significant parameter, which affects the evolution of perturbations in the cosmological perturbation theory. When the universe is assumed to be composed of a perfect fluid, $C_s^2$ is the adiabatic speed of sound defined as $C_s^2=
\frac{dp}{d\rho}$. To ensure classical stability in a system, the squared speed of sound must be positive, i.e stability occurs when $C_s^2\geq0$. This warrants that the pressure increases with the energy density. Hence, the perturbations do not grow exponentially with time. Maintaining causality requires ensuring $C_s^2$ does not exceed 1, since $C_s^2>1$ implies superluminality. Therefore, the region indicated by the inequality $0\leq C_s^2 \leq 1$ produces stable and causal solutions. These criteria can be broken in some modified gravity or non-standard fluid models, particularly those that involve Null Energy Condition (NEC) violation. Exotic matter components, including ghost fields, are frequently needed to violate NEC, which can result in both classical and quantum instabilities. Although higher-derivative scalar fields, ghost condensates, and Galileon fields are frequently employed to produce non-singular bounces, they can also induce superluminal modes or instabilities that must be properly managed. Due to the gauge-dependency of $\delta \rho$ and $\delta p$, the definition of squared speed of sound as the ratio of perturbation in pressure and density does not work in all cases. In the context of multi-fluid/field cosmology, \cite{unnikrishnan2024effective} has proposed a gauge-invariant and background-dependent definition of effective speed of sound. This effective speed simplifies to the total adiabatic sound speed in the scenario of a universe where its dynamics are influenced by multiple pure-kinetic non-canonical fields, effectively making the system akin to a system of multi-barotropic fluid on large scales. One can also refer to \cite{bertacca2008scalar, camera2011measuring,camera2011power, romano2018mess, rodrǵuez2021mess, romano2024effective} for more studies on the speed of sound.\\
In our work, since the stability analysis of the models is conducted at the background level, we have computed $c_s^2=\frac{dp}{d\rho}$, the squared speed of sound, using the reconstructed density and pressure derived for each model in the previous sections. \textcolor{black}{To obtain stable numerical derivatives, we have used a local 
polynomial (Savitzky–Golay) filter before differentiation to smooth the reconstructed 
$p_{eff}(t)$ and $\rho_{eff}(t)$. This was done only to reduce the numerical noise without changing the underlying physics of the Universe.}
 Fig. \ref{F5a} shows the resulting behaviour of the squared speed of sounds against cosmic time, for both bounce models, with the help of which, insights into the classical and causal stability of the models can be gained.  \textcolor{black}{
For clarity, we have presented four panels in Fig.~\ref{F5a}, even though the analysis pertains only to the two models we have discussed so far. The top-left and 
bottom-left panels show the trajectories for the adiabatic sound speed 
$c_s^2 = dp_{eff}/d\rho_{eff}$ for Model-I and Model-II, 
respectively. However, in the case of the viscous Van der Waals model (Model-II), the equation of 
state gives an additional microphysical speed of sound 
$c_{s, micro}^2 = \partial p_{eff}/\partial\rho_{eff}$, 
and this we have computed analytically and is shown in the top-right panel. Finally, the bottom-right 
panel exhibits the reconstructed time derivative $\dot{\rho}_{eff}$, 
just to illustrate why the adiabatic sound speed of Model-I generates a change of sign near the bounce.
Therefore, two different definitions of sound speed for Model-II, one definition for 
Model-I, and one diagnostic quantity are represented in Fig. \ref{F5a}. The region $0\leq C_s^2\leq1$ indicates stable and causal solutions since $C_s^2>1$ means superluminality.}\\

\textcolor{black}{From the figure, we note that for Model-II, the microphysical(analytic) speed of sound (top-left panel)
remains constant at $c_{s, micro}^{2} \simeq 0.55$ throughout the entire
time domain, which indicates a causally stable solution as it lies within the acceptable region.  Note that the constant value of $C_s^2$ can be attributed to the fact that $\beta\rho/\rho_c\ll1$ and $\sigma\rho/\rho_c\ll1$ for the reconstructed
$\rho_{\mathrm{eff}}(t)$ and we can expand the analytic expression to obtain
$c_{s,\mathrm{micro}}^2 \simeq \gamma + {\cal O}(\beta\rho/\rho_c,\sigma\rho/\rho_c)$.
The graph is indicative of the internal consistency of the Van der Waals fluid.
Similarly, we see that the reconstructed adiabatic sound speed (top-right panel) also remains 
positive and within the acceptable region, exhibiting a variation only at the bounce, which can be due to $\dot{\rho}_{\mathrm{eff}}$ approaching zero momentarily. Most importantly, the squared speed of sound never becomes negative, which tells us that the viscosity can act as a damping factor to prevent gradient instabilities. Thus, Model-II is observed to maintain stability throughout the bounce evolution.
}\\

\textcolor{black}{In contrast, \textbf{Model-I} (bottom-left panel) shows pathologies that are characteristic of phantom-related bounce scenarios.
The sound speed takes negative values in the contracting phase, which hints at the presence of gradient instabilities. At the bounce point, 
$c_s^2$ shows a sharp spike which is  where $\dot{\rho}_{\mathrm{eff}} = 0$ 
(bottom-right panel), indicating a stiff value of the ratio 
$\dot{p}_{\mathrm{eff}} / \dot{\rho}_{\mathrm{eff}}$.
The sound speed is also seen to be briefly exceeding unity and hence enters a transient superluminal phase.
These features are frequently associated with non-singular bouncing models involving phantom fields and higher curvature terms. While these behaviours do not automatically cause the model to be physically inconsistent, they tell us that the non-viscous model may be increasingly sensitive to the perturbations near the bounce. Several features of the graph in the lower-right panel are noteworthy. Firstly, the change of sign seen at the bounce(the shaded region). $\dot{\rho}_{eff}$
is negative during contraction, becomes 0 at $t=0$ and
positive during expansion. This zero crossing is the reason the 
adiabatic sound speed $c_s^2=\dot p_{\mathrm{eff}}/\dot\rho_{\mathrm{eff}}$ 
shows steep spikes in the bouncing phase. Secondly, we see a small dip and a peak around
$t\approx 0$, which signifies some rebalancing between geometric and field contributions as the transition occurs. Thirdly, the asymmetry and the relatively large positive values post-bounce indicate 
that the effective density grows faster than the pre-bounce decay. This may be due to an asymmetric coupling function or due to the dissipative effects of viscosity. It is important to note that this study has done the stability analysis at the preliminary level and a more detailed perturbation analysis would be required to determine the stability of the models. However, the present results do help us conclude that the presence of bulk-viscosity does lead to a more well-regulated speed of sound with no negative values while producing only a small yet physically justifiable deviation in the squared speed of sound.}

 
\section{Data Analysis}\label{6}
In this section, we scrutinise the observational viability of the bounce-inspired cosmological model developed in the previous sections. 
This is done with two aims in mind: first, to test the consistency of the scale factor that we have taken from literature, against late-time data from Type Ia supernovae, and second, to study its consequences for the early universe by computing inflationary observables while using the reconstructed potential from our model to predict its behavior in the post-bounce phase.
The late-time analysis relies on the Pantheon+SH0ES dataset \cite{Scolnic2021PantheonFull, Brout2022CosmologicalConstraints}, which sets very tight constraints on the history of background expansions, enabling us to test the bounce parametrization against observed luminosity distances.  
Both preliminary residual analysis and full Bayesian parameter estimation through Markov Chain Monte Carlo (MCMC) are performed to quantify agreement of the model with data and to put constraints on the free parameters $\{\alpha, \eta, H_0, M\}$.
The early-time calculation is done within the Gauss–Bonnet scalar field model, where the scale factor inspired by the bounce causes a reconstructed scalar potential $V(t)$ which is the same for both the viscous and non-viscous models. Using this reconstructed potential, we have computed the slow-roll parameters, the scalar spectral index $n_s$, and the tensor-to-scalar ratio $r$. 
By comparing these predictions against the Planck 2018 bounds \cite{Akrami2020PlanckInflation, aghanim2020planck}, we assess whether the model can support a viable inflationary phase consistent with observations. Collectively, these complementarity approaches provide a roundup assessment of the model at early and late times.

\subsection{Observational Viability of the Bounce-Inspired Scale Factor}
In the cosmological bouncing scenario, the scale factor is arguably the mainstay of the universe’s evolution as it gives us insights into how the universe contracts, bounces, and expands. Its shape not only determines how smooth the bounce is but also how long the contraction or expansion phases last, and how key cosmological parameters like the Hubble parameter or energy density behave. So, picking the right scale factor is crucial for making sure the evolution is non-singular and physically meaningful. A variety of its forms are available in literature, from simple power-law forms \cite{myrzakulov2014bounce,brevik2020viscous,chattopadhyay2023cosmological,yousaf2022cosmic} to more complicated ones inspired by matter \cite{brandenberger2009matter,de2015extended,raveendran2018viable,bacalhaumatter,cai2016searching}- or radiation-dominated bounces \cite{molinari2024radiation,karouby2010radiation,piazza2025time,bhattacharya2013lee}. The key is that these forms usually satisfy some important physical conditions: they’re symmetric in time around the bounce, the Hubble parameter and its derivatives stay finite, and at late times they can reproduce familiar cosmological eras.
We also want a scale factor that’s easy to work with analytically, so we can actually reconstruct scalar field potentials, calculate inflationary observables, and compare predictions with data. An apt parametrisation of scale factor gives us a neat, controllable framework where we can clearly see how changing the parameters affects the dynamics and observables. With all this in mind, we now focus on a particular scale factor, inspired by matter-bounce scenarios, which satisfies all these criteria and is also convenient enough to handle explicitly in calculations.

In order to describe a universe that avoids the initial singularity, we begin with a scale factor that naturally incorporates a smooth bounce. The form we adopted was:
\begin{equation}
a(t) = \left(\frac{\alpha}{\eta} + t^{2}\right)^{1/(2\eta)},
\end{equation}
where $\alpha$ and $\eta$ are free parameters, influencing the bouncing behaviour, to be constrained by observations. This particular choice belongs to the category of matter-bounce-inspired scale factors. What makes this class of scale factors particularly appealing is that, before the bounce, the universe behaves effectively as if it is dominated by a matter-like component ($p\approx0$). This setup naturally gives a smooth, non-singular bounce and can produce a nearly scale-invariant spectrum of perturbations. It can be noted that a model inspired by this form of $a(t)$ bounces at $t=0$ and the scale factor obtains a non-zero finite form i.e. $a(t=0)=\left(\frac{\alpha}{\eta}\right)^\frac{1}{2\eta}$. This ansatz has several appealing features. It is simple enough to handle analytically but rich enough to capture the essential physics of a bounce. Prior to this bouncing era, we can picture an early contraction followed by an expansion since the scale factor expands symmetrically on both sides(see subfig. \ref{a1} of Fig. \ref{f1}). It experiences fluctuations as it transitions smoothly from one phase to another. Under the Big Bang cosmological scenario, this transition brings about a non-singular bounce. The parameter $\alpha$ controls the minimum size at the bounce, while $\eta$ governs both the sharpness of the transition and the late-time behaviour. For a large $t$, we get $a\sim t^{\frac{1}{\eta}}$ which corresponds to a power-law form of scale factor. Interestingly, depending on the value of $\eta$, we can get standard cosmological eras like the matter-dominated or radiation-dominated eras. A more detailed description of the scale factor and the choice of $\eta$ has been given in Section \ref{1} of this study. The dependence of $a(t)$ on $t^2$ also shows us the time-reversal symmetry of the scale factor. 
The dynamics are also well-behaved. When one considers the corresponding Hubble parameter $H=\frac{\dot{a}}{a}$, one may note that it becomes 0 at $t=0$ while $\dot{H}(t)>0$ near the bouncing point. This highlights a non-singular nature of the bounce and the necessary violation of the null energy condition in this framework. Since both $H$ and $\dot H$ remain finite near the bounce, the background evolution is smooth and free of singular behaviour, consistent with the requirements of a physically reasonable bounce in Gauss--Bonnet modified gravity.
Finally, this ansatz is not only physically intuitive but also practically useful. Its analytic form makes it straightforward to reconstruct scalar field potentials and inflationary observables, and its small number of parameters allows us to constrain it directly with observational data. 
While the scale factor was originally proposed in \cite{agrawal2021matter}, our analysis provides a cross-check of its observational viability using late-time data.  This naturally leads into the next discussion on the observational viability of the bounce-inspired scale factor and, specifically, the fit to Pantheon+ supernova data.\\
\textcolor{black}{At this juncture, we would like to briefly explain why late-time observations matter in bounce cosmology. The bounce itself occurs in the early epochs of cosmological history, but we want to see if the scale factor we use here defines a single and continuous background evolution spanning from the universe’s contraction, through the bounce, and all the way into expansion. Now, the parameters $(\alpha,\eta)$ don’t just correspond to the minimum scale factor at the bounce. They also shape how the universe expands much later on, for which we have observational data—like supernova observations.
So, when we use late-time data to constrain $(\alpha,\eta)$, we’re not directly probing the bouncing era. Instead, we’re making sure the global evolution of our model is calibrated. This type of "calibration" ensures that the "post-bounce" part of the solution is observationally viable, which then imposes further constraints upon the range of parameters in which the type of early time bounce dynamics which has been discussed here is actually possible. Thus, this concept of Pantheon+ analysis can be seen as testing the consistency as well as viability of the model rather than testing the cosmological principle of bounces per se.
So, even though there’s no direct link between late-time supernova data and the bounce cosmology, this connection still matters. It’s essential for a cosmological model that describes a unified and observationally consistent cosmological framework.}

\subsection{Fit to Pantheon+ Data}
\label{sec:pantheon_fit}

To test the observational viability of the bounce-inspired scale factor in the late-time universe, we constrain its free parameters, $\alpha$ and $\eta$, along with the Hubble constant ($H_0$) and the absolute magnitude ($M$), using the Pantheon+ Type Ia supernova (SN Ia) data set. \textcolor{black}{We note that for this section, we exclude Model-II i.e. the bulk viscosity effect. The framework for the second model uses bulk viscosity as a tool to control dynamic behavior during the bouncing period while simultaneously suppressing anisotropies ensuring a non-singular bounce. The effective pressure receives its viscous contribution through terms proportional to $\xi(H)$) and the chosen parameterization $\xi(H)=\tau (3H)^m$ (with $m=2$) which means that the viscous effects scale as higher powers of the Hubble parameter. Thus, the rapid changes in the Hubble parameter during the bounce period create significant effects which will decay rapidly during the subsequent expansion phase. The viscosity-induced terms become extremely small in late times, which are important for SN\,Ia observations and they do not have any significant impact on the background expansion history. Consequently, the Pantheon+ dataset cannot constrain bulk-viscous parameters which leads to their exclusion from our MCMC analysis because the analysis requires exclusive focus on parameters that control late-time expansion. The observational viability of post-bounce evolution can be evaluated through this separation while maintaining viscosity as a mechanism pertaining only to the early-time dynamics.}
\textcolor{black}{At this juncture of our study, we note that although the bouncing behaviour is a feature of the early universe and the cosmological Pantheon+ dataset is categorised as late-time data, the parametrisation of the scale factor we have used in this study is not restricted to that era only. The scale factor $a(t)$ remains mathematically valid and continuous throughout all the cosmic times, offering a single smooth background evolution that links the stages of contraction, bounce, and expansion. We particularly point out that the parameters $\alpha$ and $\eta$ determine the minimal scale factor at the bounce ($a_{min}$) and the late-time asymptotic behavior, where $a(t)\propto t^{\frac{1}{\eta}}$ and the corresponding Hubble evolution can be directly compared with observations. Hence, by constraining these parameters using late-time data we do not intend to scrutinize just the bounce itself, but it serves two major purposes: (i) it verifies the observational validity of the post-bounce expansion proposed by the model, making sure that the late-time limit does not disagree with the current expansion–history measurements, and (ii) it offers a data-based calibration of the parameter space that is afterwards used to investigate the early-time bounce dynamics.
This approach is motivated by the fact that a cosmological model that asserts it can describe the entire evolution of the universe should consistently incorporate both the early epoch's theoretical expectations and the late-time observational constraints. If not, the chosen scale factor would remain to be a mere theory. We have tried to establish a continuous evolutionary framework preliminarily by fixing
($\alpha,\eta$) via the Pantheon+ dataset and then applying the same best-fit values to inquire into the bounce region: the same parameterisation rules the contracting phase, the bounce, and the late-time acceleration. Such an approach has been taken in the bouncing and emergent universe scenarios through parametrisation, where the late-time cosmological data serves not only to anchor the free parameters but also to increase the physical credibility of the model. From this perspective, the Pantheon+ analysis works as an observational consistency check rather than a direct probe into the bounce, which in turn makes the scenario developed here overall more robust.}

\subsubsection{Methodology: Distance Modulus and Likelihood}

The primary observable in SN Ia cosmology is the distance modulus, $\mu(z)$, which relates the observed apparent magnitude ($m_B$) to the absolute magnitude ($M$) of the supernova:
\begin{equation}
\mu(z) = m_B - M = 5 \log_{10}\left(\frac{D_L(z)}{\SI{10}{pc}}\right),
\label{eq:mu}
\end{equation}
where $D_L(z)$ is the luminosity distance. Assuming a flat universe, the luminosity distance is related to the Hubble parameter $H(z)$ by:
\begin{equation}
D_L(z) = (1+z) \frac{c}{H_0} \int_0^z \frac{dz'}{E(z')},
\label{eq:dl}
\end{equation}
where $c$ is the speed of light, $H_0$ is the Hubble constant, and $E(z) = H(z)/H_0$ is the dimensionless Hubble function.

The specific dimensionless Hubble function corresponding to our bounce-inspired model, as parameterized for late-time analysis, is given by:
\begin{equation}
E(z; \alpha, \eta) = \sqrt{\alpha (1+z)^{3\eta} + (1-\alpha)}.
\label{eq:ez}
\end{equation}
The term $\alpha$ controls the influence of the matter-like bounce component, while $\eta$ dictates the effective equation of state at late times.

The model parameters are constrained using the standard chi-squared likelihood function for supernova data. For a set of observed distance moduli $\mu_{\text{obs},i}$ with uncertainties $\sigma_{\mu,i}$, the $\chi^2$ statistic is:
\begin{equation}
\chi^2(\theta) = \sum_{i} \left[\frac{\mu_{\text{model}}(z_i; \theta) - \mu_{\text{obs},i}}{\sigma_{\mu,i}}\right]^2,
\end{equation}
where $\theta = \{\alpha, \eta, H_0, M\}$ is the set of parameters. The log-likelihood function is then defined as $\ln \mathcal{L}(\theta) \propto -\frac{1}{2} \chi^2(\theta)$.

\subsubsection{MCMC Implementation and Results}

\textcolor{black}{We employed a Markov Chain Monte Carlo (MCMC) analysis using the \texttt{emcee} sampler with $\num{48}$ walkers. A total of $\mathbf{150\,000}$ sampling steps were executed, with a burn-in of $\mathbf{6000}$ steps and a thinning factor of $\mathbf{5}$, resulting in approximately $\mathbf{1.39\times10^{6}}$ posterior samples.Uniform priors were adopted:
\[
\alpha \in [10^{-6},10^{4}],\qquad
\eta \in [0.01,5.0],\qquad
H_0 \in [10,120],\qquad
M\in[-25,0].
\]}
\textcolor{black}{To accelerate the likelihood evaluation over the Pantheon+ dataset, the luminosity distance integral was computed using a vectorized trapezoidal integration scheme on a fixed redshift grid, combined with a cache keyed to $(\alpha,\eta,H_0)$ to avoid redundant recomputation. This significantly improves efficiency for long MCMC chains.}

\begin{figure}[H]
    \centering
    \includegraphics[width=0.9\linewidth]{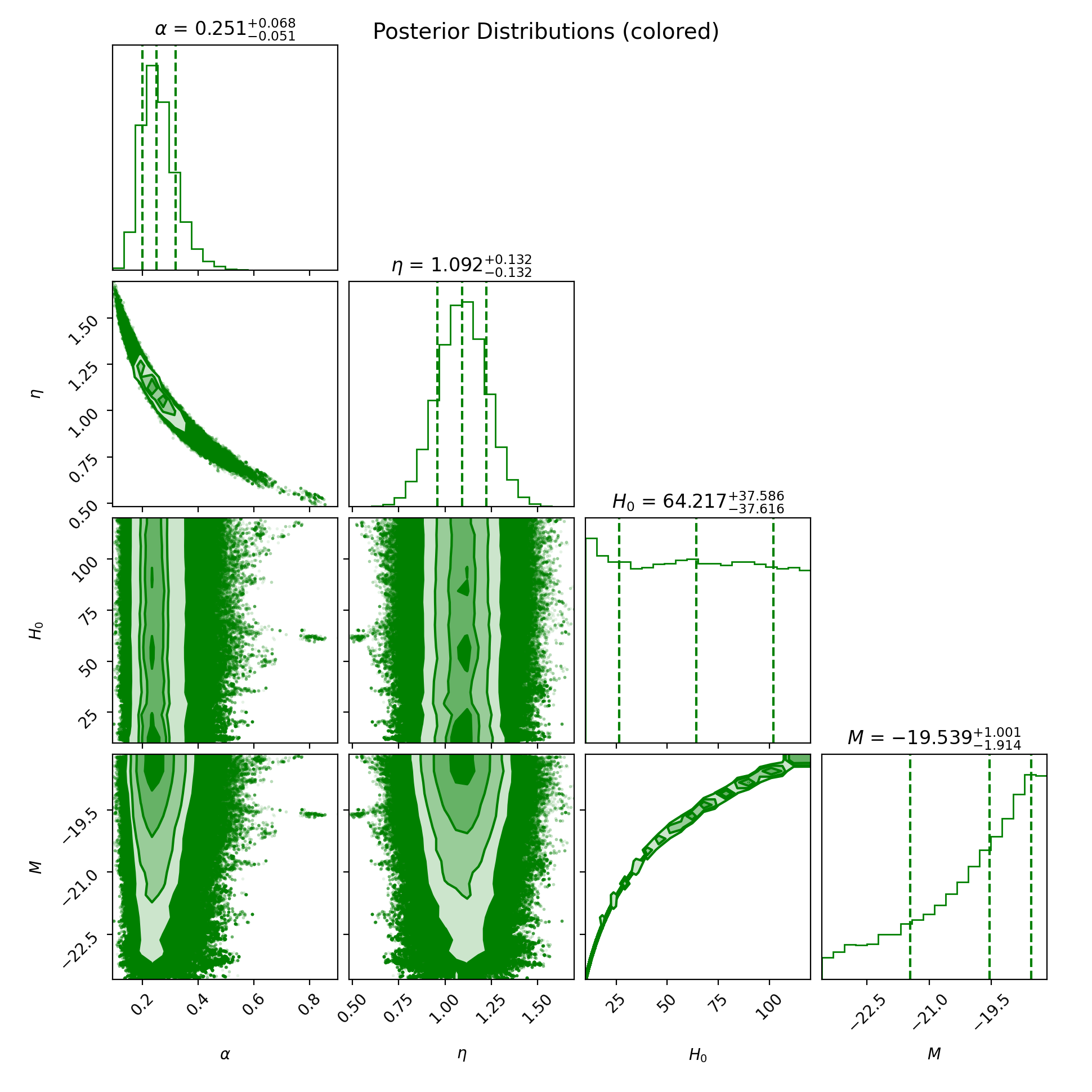}
    \caption{\textcolor{black}{MCMC Constraints and Degeneracies from Pantheon+ Data. Marginalized one- and two-dimensional posterior distributions for the model parameters ($\alpha$, $\eta$, $H_0$, $M$) derived from the MCMC chain. The diagonal panels show the one-dimensional posterior histograms, with the median (solid line) and $68\%$ confidence intervals, indicated by dashed lines. The off-diagonal panels display the joint $1\sigma$ and $2\sigma$ contours, highlighting the strong linear degeneracy between $H_0$ and $M$, and the non-linear correlation between the core model parameters $\alpha$ and $\eta$.}}
    \label{pcc}
\end{figure}

\textcolor{black}{Convergence was assessed using ArviZ diagnostics. The core model parameters $(\alpha,\eta)$ show R-hat values extremely close to unity and large effective sample sizes, indicating excellent convergence. In contrast, the parameters $(H_0,M)$ exhibit mildly elevated R-hat values ($\sim 1.014$), arising from the known SN-only degeneracy between the absolute magnitude and the Hubble constant. This behaviour is expected and does not affect the robustness of the constraints on $(\alpha,\eta)$.
The one- and two-dimensional marginalized posterior distributions are summarized in the corner plot (Fig. \ref{pcc}), and the median parameter estimates with their $68\%$ confidence intervals are presented in Table \ref{tab:results}.The reduced chi-squared is computed using $N_{\rm data}-4$ degrees of freedom, with $N_{\rm data}=1701$ for the Pantheon+ sample.}

\begin{table}[H]
\centering
\caption{MCMC Constraints for the Bounce Model Parameters from Pantheon+ Data (Median $\pm 1\sigma$).}
\label{tab:results}
\begin{tabular}{lc}
\toprule
Parameter & Constraint ($\text{Median} \pm 1\sigma$) \\
\midrule
$\alpha$ & $\alphafit \pm \num{0.0595}$ \\
$\eta$ & $\etafit \pm \num{0.1321}$ \\
$H_0$ ($\si{km.s^{-1}.Mpc^{-1}}$) & $\Hnaughtfit \pm \num{37.601}$ \\
$M$ (mag) & $\Mfit \, ^{+1.001}_{-1.914}$ \\
\bottomrule
\end{tabular}
\end{table}

\begin{figure}[H]
\centering
    
   \includegraphics[width=0.6\linewidth]{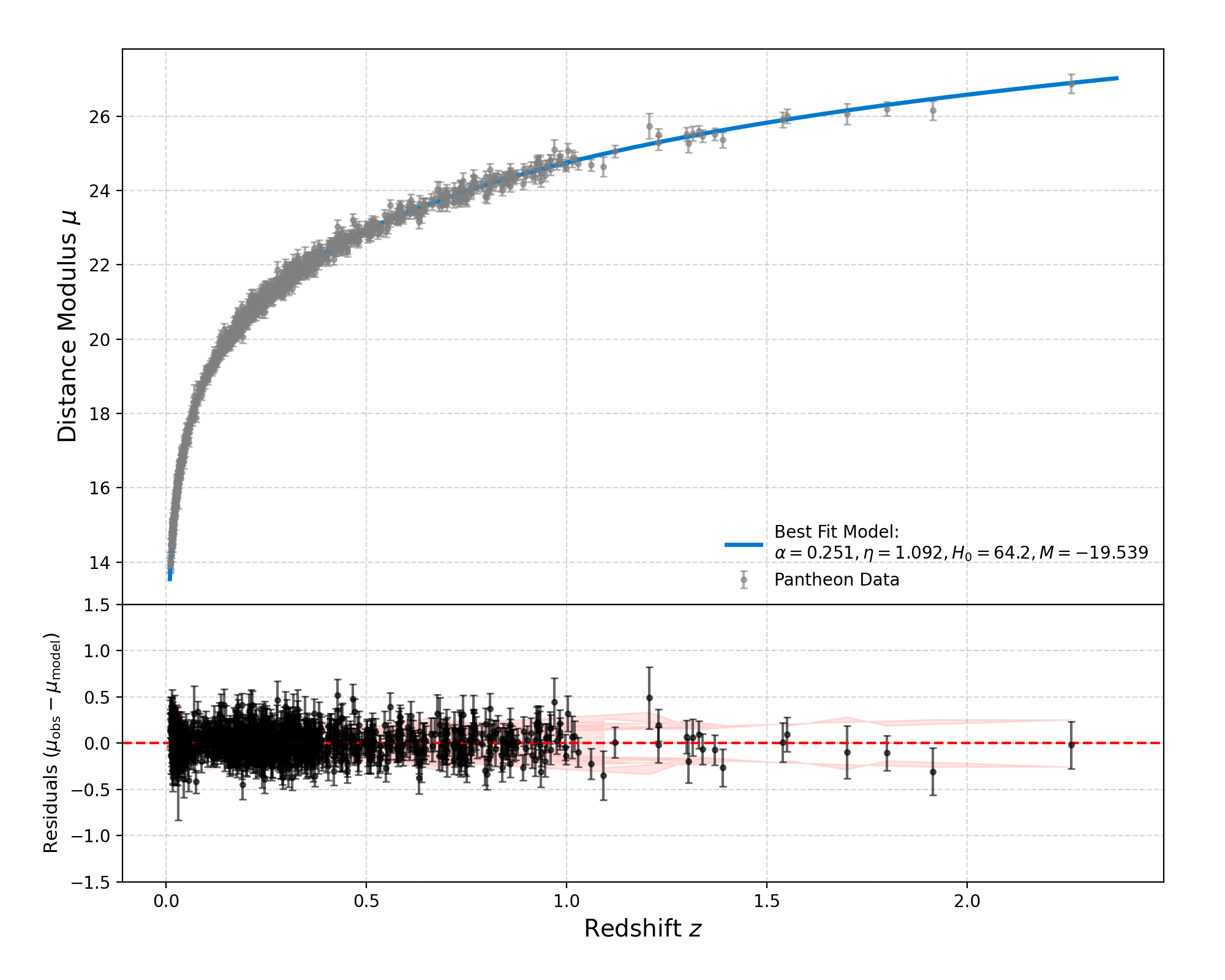}
 \caption{\textcolor{black}{Distance Modulus Hubble diagram for the Pantheon+ SNe Ia sample fitted with the modified expansion model. The blue curve represents the best-fit parameters ($\alpha=\alphafit$, $\eta=\etafit$, $H_0=\Hnaughtfit~\mathrm{km\,s^{-1}\,Mpc^{-1}}$, $M=\Mfit$), while the gray points with error bars denote the observed distance moduli. The lower panel shows the residuals, which scatter symmetrically around zero, indicating that the model provides an excellent fit with $\chi^2_{\mathrm{min}}=\mathbf{\chiSqMin}$ and $\chi^2_{\mathrm{red}}=\mathbf{\chiSqRed}$.}}
\label{fig:hubble_diagram}
\end{figure}

\subsubsection{Interpretation of Results}

\textcolor{black}{The MCMC analysis yields a minimum chi-squared of $\mathbf{\chiSqMin}$ for $\mathbf{\DoF}$ degrees of freedom, giving a reduced chi-squared of $\chi^2_{\rm red} = \mathbf{\chiSqRed}$. Since $\chi^2_{\rm red} \approx 1$, the model provides an excellent statistical fit to the Pantheon+ observations.}

\textcolor{black}{The best-fit distance modulus curve in Fig.~\ref{fig:hubble_diagram} passes centrally through the SN Ia data, and the residuals show no significant systematic trends, confirming the observational viability of the bounce-inspired parametrization.}

\textcolor{black}{The two core model parameters are constrained as follows:
\begin{itemize}
\item $\alpha \approx \alphafit$ indicates that the bounce-related component contributes non-trivially but remains subdominant at late times.
\item $\eta \approx \etafit$ determines the effective time–dependence of the late-time scale factor. \textcolor{black}{Values of $\eta$ differing from unity imply small deviations from standard $\Lambda$CDM-like scaling while remaining compatible with current SN-only constraints.}
\end{itemize}
Figure~\ref{fig:hubble_diagram} illustrates the performance of the bounce-inspired
model against the Pantheon+ SN\,Ia sample through the Hubble diagram and its 
corresponding residuals. The upper panel shows that the best-fit theoretical 
distance modulus curve, computed using the median MCMC parameters, traces the 
observed luminosity–distance relation remarkably well across the full redshift 
range. The model follows the sharp low-$z$ rise and the gradual high-$z$ 
flattening of the data without exhibiting any systematic departures. This visual 
agreement is quantitatively supported by the reduced chi-squared value 
$\chi^2_{\mathrm{red}}\approx\chiSqRed$, indicating statistical consistency with 
the dataset.
The residual panel further reinforces this conclusion. The points scatter around 
zero with no discernible redshift-dependent trend or coherent structure, implying 
that the model captures the overall shape of the expansion history encoded in the 
Pantheon+ compilation. The residuals remain well within the $1\sigma$ envelope 
throughout, including the transition region near $z\simeq 1$ where many modified 
expansion histories diverge from $\Lambda$CDM-like behaviour. The absence of 
systematic drift in the residuals suggests that the bounce-motivated 
parametrization offers a stable and unbiased description of the luminosity 
distance, demonstrating that the late-time limit of the scale factor used in this
work is fully compatible with observational constraints from SN\,Ia data alone.}

\textcolor{black}{\textbf{Corner Plot Interpretation.} } \\
\textcolor{black}{Figure~\ref{pcc} shows:
\begin{itemize}
    \item well-peaked, tightly constrained posteriors for $(\alpha,\eta)$,
    \item \textcolor{black}{broad, elongated posteriors for $(H_0,M)$ due to their intrinsic SN-only degeneracy},
    \item a strong, nearly linear degeneracy between $\alpha$ and $\eta$(negative correlation) reflecting their shared role in shaping $E(z)$,
    \item the expected strong, curved degeneracy (often called the banana-degeneracy) between $H_0$ and $M$, which is characteristic of the supernova distance modulus calculation. This degeneracy is the primary reason for the broad 1D posteriors of $H_0$ and $M$.
\end{itemize}}

\textcolor{black}{The main results are not actually affected by degeneracies. The strong $H_0-M$ degeneracy just shows the well-known fact that SN\,Ia data alone can’t pin down the absolute distance scale, so this mostly tweaks nuisance parameters, not the real expansion dynamics. The $\alpha$ and $\eta$ correlation is a similar story—it’s a degeneracy corresponding to expansion histories that look almost the same in terms of luminosity–distance at late times. Still, the data constrain these parameters to a narrow parameter space, so the bounce-inspired scale factor stays robust, both qualitatively and quantitatively, at late times. To really solve these degeneracies would require more datasets like BAO or CMB, which we leave for future work. In short, the constraints on $(\alpha,\eta)$ are robust and show excellent convergence.}

Thus, the bounce-inspired scale factor is statistically consistent with late-time supernova observations, and the Pantheon+ dataset strongly constrains the parameters governing its late-time behaviour.

\subsection{Early-Time Slow-Roll Dynamics and Inflationary Viability}\label{sub1}
Matter-bounce models offer a very intuitive and physically rich framework to explore the pre-inflationary universe. They also allow us to compute key inflationary observables directly from the dynamics, such as the scalar spectral index $n_s$ and the tensor-to-scalar ratio $r$.
 What is even more interesting is that, with an appropriate choice of scalar field dynamics or modified gravity contributions—such as the Gauss-Bonnet coupling—the universe can transition seamlessly after the bounce into a period of accelerated expansion, almost like slipping into a slow-roll inflationary phase.
 Thus, to check if our bounce model can give a viable inflationary phase after the bouncing epoch, we have done a detailed numerical implementation of the slow-roll parameters derived from the reconstructed scalar potential $V(t)$ as given in Eq. \ref{V1} within the Gauss-Bonnet phantom scalar field framework. This adds to the late-time observations validation from the Pantheon+ dataset and investigates the early post-bounce regime where inflationary dynamics could come into play. In the subsequent sections, we apply this to compute inflationary observables, namely the scalar spectral index $n_s$ and tensor-to-scalar ratio $r$, which are then confronted with the CMB data from Planck 2018.\\
 \textcolor{black}{We bring to the reader's attention that the slow–roll analysis done in this study is not based on the assumption that primordial fluctuations take place during the bouncing epoch. Rather, the $V(t)$ potential that is reconstructed through the Gauss–Bonnet scalar field is utilized for probing the post-bounce dynamics. The nature of many bouncing and emergent scenarios is such that the bounce just serves as a pre-inflationary mechanism that regularises the initial conditions and then a slow–roll phase further growing to generate the observable perturbation spectra. Therefore, the goal of this section is to see if the era after the bounce could be the same theoretical structure which produced the non-singular bounce that naturally leads to a slow-roll regime producing viable inflationary observables.
So, the calculation of slow–roll parameters $(\epsilon,\eta)$ and the related observables $(n_s,r)$ should be regarded as an inflationary viability test of the post-bounce branch of the model rather than as a perturbative description of the bounce itself. The potential reconstructed is computed only during the expanding phase($t>0$) where standard slow–roll approximations are valid which ensures that there is no conceptual inconsistency. Through this method, we can determine if the model can both offer an early-time evolution that is non-singular and an inflationary phase that is consistent with observation, thus establishing a link between the bounce dynamics and the inflationary paradigm in a unified framework.}

\paragraph{Sampling Strategy and Ensemble Construction.}
\textcolor{black}{Using a Monte Carlo method, we produce an assortment of inflationary trajectories by exploring the entire parameter space of the reconstructed Gauss-Bonnet model potential $V(t)$ through a random sampling process. An advanced search method is used, instead of the sampling of Pantheon+ best-fits, which is combined with fixed coupling parameters. For this investigation, the Gauss-Bonnet coupling parameters $\alpha$ and $\eta$ are set at $\alpha = 0.2506$ and $\eta = 1.0924$, respectively.} The other parameters $\{n, \phi_0, C_1, f_0, \rho_{m_0}, \lambda, B\}$ are uniformly distributed according to the ranges given in Table~\ref{tab:params_new}, thus taking a long way through the parameter space. \textcolor{black}{The model is tested in the time domain $t \in [5.0, 500.0]$, which is selected to include the entire slow-roll period.}

\begin{table}[!ht]
\centering
\caption{\textcolor{black}{Parameter ranges and sampling types used for the Monte Carlo ensemble scan of the $n_s$-$r$ plane.}}
\arrayrulecolor{blue}
\begin{tabular}{|c|c|c|c|}
\hline
\textcolor{black}{\textbf{Parameter}} & \textcolor{black}{\textbf{Symbol}} & \textcolor{black}{\textbf{Sampling Range}} & \textcolor{black}{\textbf{Sampling Type}} \\
\hline
\textcolor{black}{Model exponent} & \textcolor{black}{$n$} & \textcolor{black}{$[1.2, 4.0]$} & \textcolor{black}{Uniform} \\
\textcolor{black}{Scalar field constant} & \textcolor{black}{$\phi_0$} & \textcolor{black}{$[5\times10^{-4}, 2\times10^{-3}]$} & \textcolor{black}{Uniform} \\
\textcolor{black}{Integration constant} & \textcolor{black}{$C_1$} & \textcolor{black}{$[5\times10^{-4}, 1\times10^{-2}]$} & \textcolor{black}{Uniform} \\
\textcolor{black}{Coupling constant} & \textcolor{black}{$f_0$} & \textcolor{black}{$[10^{-5}, 10^{-3}]$} & \textcolor{black}{Uniform} \\
\textcolor{black}{Matter density parameter} & \textcolor{black}{$\rho_{m_0}$} & \textcolor{black}{$[0.3, 0.8]$} & \textcolor{black}{Uniform} \\
\textcolor{black}{Model coefficient} & \textcolor{black}{$\lambda$} & \textcolor{black}{$[10^{-4}, 10^{-2}]$} & \textcolor{black}{Uniform} \\
\textcolor{black}{Model coefficient} & \textcolor{black}{$B$} & \textcolor{black}{$[10^{-5}, 10^{-2}]$} & \textcolor{black}{Uniform} \\
\hline
\end{tabular}
\label{tab:params_new}
\end{table}

\paragraph{\textcolor{black}{Numerical Evaluation of Observables and Cost Function.}}
\textcolor{black}{The numerical evaluation of the scalar potential $V(t)$ is done for each set of sampled parameters in the range of $t \in [5.0, 500.0]$. After that, the derivatives $\dot{V}(t)$ and $\ddot{V}(t)$ are then computed through finite differencing using the high-precision $\Delta t$ equivalent, which is a part of the numerical library ($\texttt{numpy.gradient}$ on $t_{vals}$).} The standard slow-roll parameters are defined as:
\begin{align}
\epsilon(t) &= \frac{1}{2} \left( \frac{\dot{V}(t)}{V(t)} \right)^2, \\
\eta(t) &= \frac{\ddot{V}(t)}{V(t)},
\end{align}
\textcolor{black}{To obtain a single point for the $n_s$--$r$ plane, we compute the mean values over the time domain $t \in [5.0, 500.0]$:}
\begin{align}
n_s &= \left< 1 - 6\epsilon(t) + 2\eta(t) \right>_t, \\
r &= \left< 16\epsilon(t) \right>_t.
\end{align}
\textcolor{black}{We used a Cost Function to quantify the viability of each model point $(\mathbf{n_s}, \mathbf{r})$. This measures its deviation from the Planck 2018 central value ($\mathbf{n_{s, \text{central}}} = 0.965$, $\mathbf{n_{s, \sigma}} = 0.004$) and the upper bound ($\mathbf{r_{\text{upper}}} = 0.07$):}
$$
\text{Cost} = \left( \frac{n_s - n_{s, \text{central}}}{n_{s, \sigma}} \right)^2 + \left( \frac{r}{r_{\text{upper}}} \right)^2.
$$

\textcolor{black}{We use the ensemble scan to identify the parameter region that gives the best-fit observables, rather than to perform a statistical distribution analysis. Thus, we emphasize that the ensemble sampling is employed purely as an optimization tool to locate viable configurations in the high-dimensional parameter space, rather than to infer statistical posterior distributions.}

\begin{figure}[!ht]
\includegraphics[width=\linewidth]{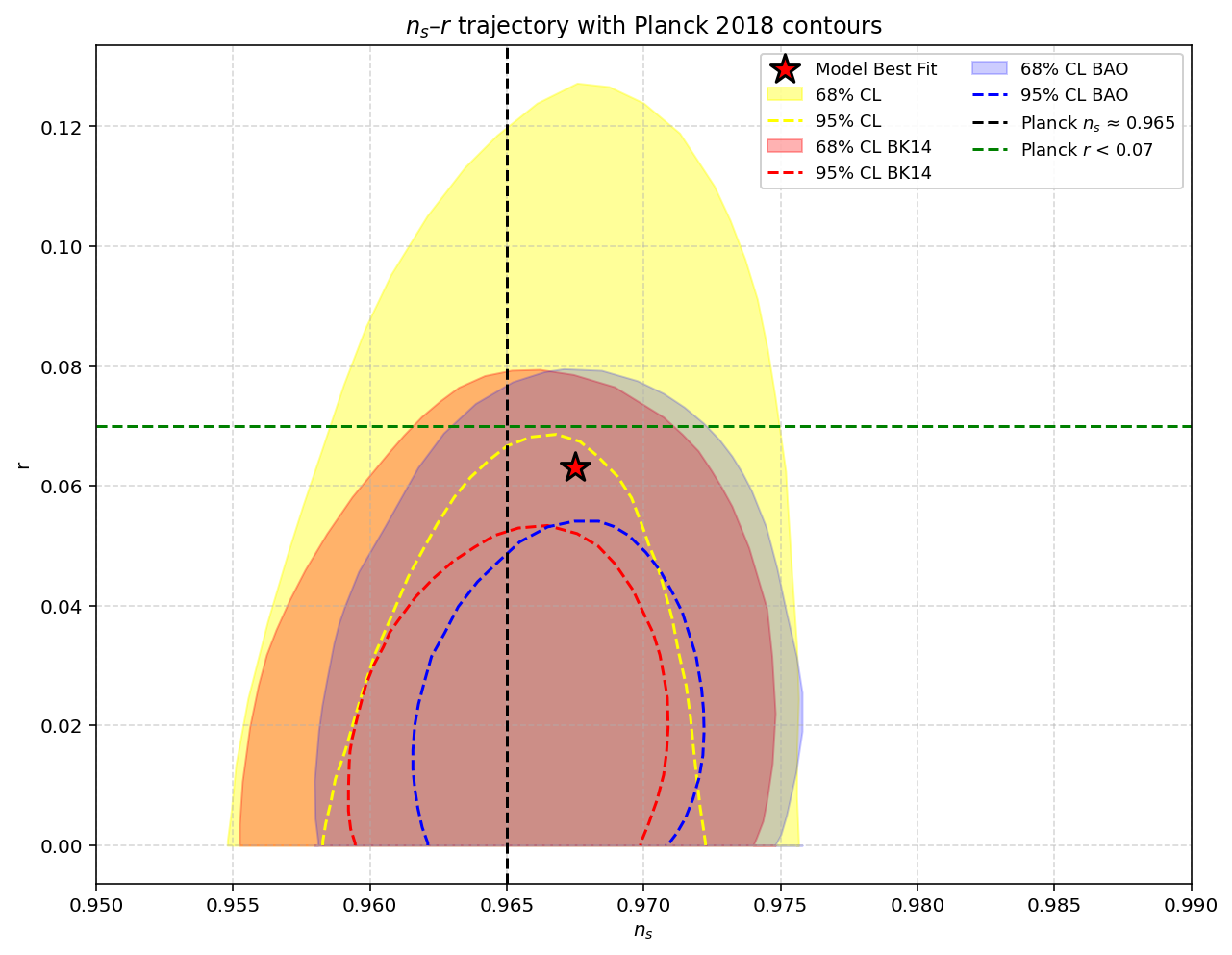}
    \caption{\textcolor{black}{The scalar spectral index $n_s$ and the tensor-to-scalar ratio $r$ value predicted by the Best-Fit Model Prediction
    , which is denoted by the red {$\star$}.
    Colored contours represent $68\%$ and $95\%$ confidence level (CL) observational bounds.
    Reference lines are at $n_s \approx 0.965$ (dashed black) and $r < 0.07$ (dashed green)}}\label{nsr}
\end{figure}


\textcolor{black}{In the plane of $n_s$ and $r$, the figure \ref{nsr} shows the position of the best-fit prediction made by the model in relation to the confidence regions of Planck 2018, BAO, and BK14. The model leads to a scalar spectral index of $n_s \approx 0.967$, which is inside the $68\%$ confidence region of the Planck constraints, thus making a case for a red-tilted spectrum. The corresponding tensor-to-scalar ratio of $r \approx 0.063$ is still less than the present upper limit of $r < 0.07$, hence the best-fit point is positioned within the range allowed by observations.}

\textcolor{black}{The interpretation of the plot is based on the fact that it exhibits only one best-fit prediction, and not a complete sampling of the parameter space. The bounce-inspired scale factor incorporated into the Gauss–Bonnet scalar field framework is shown to be able to generate inflationary observables that are consistent with present-day CMB constraints. This association reinforces the case for a non-singular bounce that is followed naturally by a slow-roll inflation phase, thus connecting early-universe dynamics within a single description.}

        \label{fig:nsr_trajectory}

\section{Conclusion}\label{7}
Among numerous ideas, the theory of bounce in cosmology has played the role of a promising alternative or an imposing challenge to the inflationary paradigm. Much work has been done in recent years to create singular, nonsingular, and well-behaved bouncing models. Motivated by this framework, we have investigated cosmological bounce driven by phantom scalar field with the Gauss-Bonnet term using a specific ansatz for the scale factor $\alpha=\left(\frac{\alpha}{\eta}+t^2\right)^\frac{1}{2\eta}$ which describes a non-singular bounce. Utilising the field equations pertaining to the Einstein Gauss-Bonnet scalar field, we have explored two different cases/ models: Model-I without viscosity and Model-II with viscosity within the same scalar field framework. To facilitate analytic expressions for the pressure and energy density in the case of both models, the scalar field $\phi=\phi_0 t^n$  was selected with the functional form of $f=f_0 \phi=f_0\phi_0 t^n$ and a coupling function $g(\phi)=Be^{\lambda\phi_{0}t^{n}}$. The reconstructed potential $V(t)$, obtained using the best-fit parameters, in the specified field was plotted against cosmic time as shown in Fig. (\ref{F1a}).  \textcolor{black}{It displays a slightly asymmetric behaviour around the bounce with its minimum around $t\approx0$ at both sides of the bouncing point. This highlights some imbalance during the transitioning phase. It is important to note that the curve is free of singularities, which underpins the non-singular transition from contraction to expansion.} This potential, shaped by hypergeometric contributions from the Gauss–Bonnet coupling, was then used as the basis for connecting the bounce framework to slow-roll analysis and the resulting predictions in the $n_s$--$r$ plane, thereby bridging theoretical reconstruction with observational consistency. In addition to this, the resulting reconstructed density, pressure and EoS parameter were analysed through the plots as depicted in \ref{F1b}. \textcolor{black}{The reconstructed energy density also shows an asymmetric behaviour, starting at highly positive values in the pre-bounce era and then decreasing monotonically. It reaches a shallow minimum at around $t\approx-1$ and then was observed to cross 0 at the bouncing point, after which the density values continue to increase in the expanding phase. It should be noted that the negative values of density for a momentary period are expected within the context of fluids that are reconstructed in the Gauss-Bonnet fields. The pressure $p_eff(t)$, on the other hand, behaves in a complementary manner to the density. While it is asymmetric, the pressure begins at highly negative values. Similar to $\rho_{eff}(t)$, the pressure crosses 0 at $t=0$. Following this, the pressure becomes rapidly negative which means that the strong epulsive gravity can be induced for post-bounce accelerated expansion in the early universe. The reconstructed EoS parameter $\omega(t)$ exhibits an oscillatory behaviour. The EoS parameter remains deeply negative in the pre-bounce phase and dips a little at around $t\approx-0.75$. Then it begins to rise steadily and crosses the phantom divide. After $t\approx-0.5$, the EoS shows a quintessence behaviour and crosses the pressureless state($\omega(t)=0$) at the bouncing point. However, the EoS parameter immediately begins to decrease and crosses the phantom divide once again at around $t\approx0.5$. This transition into the phantom regime occurs smoothly and continues to remain in this regime throughout. This oscillatory behaviour is indicative of a dynamic dark-energy-like behaviour of the fluid that is driven by the dynamics between the scalar-field coupling and the Gauss-Bonnet term. The viscous model does not present this oscillatory behaviour. Instead, as seen in Fig. \ref{eosv}, the EoS parameter for the viscous model shows violent variations around the bouncing point. In the contracting phase, the curve begins at the pressurless state ($\omega(t)=0$) and starts rising around $t\approx-2$. Two vertical singularities are also observed, showing divergences towards $\pm \infty$, in both the pre-bounce and post-bounce eras. Another point of note we make here is that around $t=0$, the EoS behaviour is finitely oscillatory. Despite the divergences and the oscillations, we observe that the EoS shows quintessence behaviour throughout. Even so, the divergence is purely due to the vanishing of $\rho_{eff}$ as seen in Fig. \ref{rho}. Comparing the behaviour of the EoS parameters of the two models, we can conclude that the bulk viscosity affects the effective dynamics, which leads to avoidance of phantom and prolonged oscillatory behaviours for a relatively more symmetric and well-behaved evolution.}\\
We then looked at the standard energy conditions to further evaluate the physical plausibility of these scenarios. For the non-viscous model, the NEC and SEC remain violated throughout the time domain, as can be seen in subfigures \ref{n1} of Fig. \ref{F3a}, which is required for obtaining bounce in scalar-field cosmology.  \textcolor{black}{Notably, both the NEC and SEC reach their minimum value at $t=0$. The NEC reaches zero value briefly before and after the bounce, but the violation of the NEC is characteristic of phantom-driven bounce models. The continuous negative violations of the SEC highlight the repulsive gravity generated by the phantom field and Gauss-Bonnet coupling. However, the DEC is obeyed in both the expanding and contracting phases. This indicates the dominance of energy density over pressure throughout. Moreover, it highlights the point that the fluid respects causal behaviour. This supports the physical viability of the model within the scalar field framework.}
In contrast, a graphical analysis of the energy conditions for the viscous model (subfigure \ref{s1} of Fig. \ref{F3a})shows that the \textcolor{black}{NEC, SEC and DEC remain positive throughout the cosmic evolution due to the presence of bulk viscosity, except for a brief period around the bounce where all three dip below zero. Thus, this combination of mostly persistent obeyance of energy conditions emphasises the significant impact of viscosity on the effective fluid wherein the violations are confined to the region around the bounce and the energy conditions are satisfied once again in post-bounce era, ensuring a relatively more controlled evolution, which was not the case for the non-viscous model.} \\
In addition to the study of energy conditions, it is imperative to investigate the dynamical stability of both models to assess their feasibility even further. The squared speed of sound, $C_s^2$, which tracks the evolution of perturbations in the cosmic fluid with time, is a commonly utilised criterion for such an analysis. A positive $C_s^2$ is an indication of classical stability against small perturbations. Thus, a stability analysis at the background level was carried out in the last phase of our study, \textcolor{black}{which further strengthens the difference between our two bounce models. The graphs (Fig. \ref{F5a}) indicate that while the non-viscous model shows sharp divergences near the bounce and is negative throughout the time domain, reflecting the typical characteristics associated with phantom-driven evolution, the viscous model shows us that the analytically derived microphysical sound speed as well as the adiabatic sound speed stays in the causal region $0\leq C_s^{2}\leq1$. This shows us that bulk viscosity regularises the dynamics while suppressing instabilities and ensuring causal and classical stability of the cosmic fluid throughout the bounce.
Thus, when we compare the viscous and non-viscous bouncing cosmologies in the Gauss-Bonnet scalar field framework, we observe that though both models exhibit a non-singular bounce, viscosity does bring about a significant difference in stability behaviour and cosmological dynamics in the post-bounce regime.}
Beyond theoretical consistency, we have also assessed the observational viability of the proposed bouncing scale factor. A Bayesian MCMC analysis using Pantheon+ supernovae data was performed, with posterior corner plots providing constraints on the free parameters $\alpha$, $\eta$, and $H_0$. These plots not only demonstrate correlations among parameters but also confirm that the best-fit values lie well within the observationally allowed regions. To further test the robustness of the fit, residuals were analysed, showing a symmetric scatter about zero with no systematic deviation. The reduced chi-squared value of \textcolor{black}{$\chi^2_{red}$ = 0.995} further reinforces the compatibility of the bounce-inspired model with observational data, indicating that the model is not only mathematically consistent but also statistically viable. In addition, inflationary observables were reconstructed from the scalar field potential, with the scalar spectral index $n_s$ and the tensor-to-scalar ratio $r$ evaluated for the best-fit parameter. \textcolor{black}{The best-fit prediction gives $n_s \approx 0.967$ and $r \approx 0.063$, which places the model within the $68\%$ confidence region of the Planck 2018 confidence contours in the $n_s$--$r$ plane, supporting the possibility that the bounce scenario can support a slow-roll inflationary phase after the bounce. An important outcome of our analysis concerns the slight deviation of $n_s$ from the Planck central value of $0.965$ and a relatively high value of $r$, which we acknowledge, can be seen as minor tensions and hints at the potential limits of the current parametrisations.}However, such a feature is not uncommon in bouncing cosmologies, where additional mechanisms (viscous effects, modified couplings, or alternative potentials) may be required to suppress tensor modes. In this sense, the elevated predictions for $r$ \textcolor{black}{and the deviated values of $n_s$} can be interpreted as a direction for refinement, pointing toward possible extensions of the model that reconcile it more fully with present CMB bounds. \textcolor{black}{Nonetheless, the model is broadly compatible with the observations and demonstrates that the scenario is both theoretically interesting and observationally testable, motivating further investigation.}

Although our models demonstrate theoretical and observational consistency, our study has certain limitations. Firstly, our reconstruction scheme relies on ansätze for the scale factor, scalar field($\phi$) and coupling functions. This can restrict the generality of the results obtained. Thus, one can also point out that the introduction of a phantom scalar field and Gauss-Bonnet term, while capable of producing a smooth bounce, may lead to instabilities or unconventional behaviour in a more general setting. Furthermore, we have seen how the viscous model produces consistent NEC and SEC violations with a highly negative EoS parameter. This could lead to uncertainties in the late-time epoch of the universe, specifically to future singularities like the Big Rip. The higher tensor-to-scalar ratio than observational bounds is another limitation which can be dealt with by employing additional mechanisms that could suppress tensor modes. Furthermore, our model primarily deals with Pantheon+ and Planck2018 datasets. A well-rounded comparison with other myriad datasets available could lead to a more complete validation of the model.
Building upon this study, future research could focus on extending the analysis to include additional observational datasets such as baryon acoustic oscillations (BAO), cosmic chronometers, and gravitational wave signals \cite{dhankar2025testing} that would provide a wide view of cosmological parameters and the dynamics of the universe. On the other hand, the search for signatures of this scenario in these datasets could help understand the early universe better and test the feasibility of such models \cite{li2024primordial}. 
\textcolor{black}{Last but not the least, our analysis recognizes a limitation with its dependent dynamical and early-universe investigations, which only employed parameter values ($\alpha,\eta$) from the Pantheon+ dataset that were the best-fits. We  have not performed a full distribution of the MCMC posterior into the bounce dynamics and the computation of inflationary observables in this work. Thus, at this point, the parameter uncertainties on the bounce characteristics and the predicted values of $(n_s,r)$ cannot be quantified.
Thus, a natural continuation of this research would involve a comprehensive Bayesian analysis where the posterior samples are transferred to the early-time regime, and this way, the bounce characteristics and inflationary predictions would be provided with credible intervals, and also broader cosmological model comparison would be allowed.
}

\appendix\label{a}
\section{\textcolor{black}{Appendix}}
\subsection{\textcolor{black}{Derivation of \texorpdfstring{$\mathrm{FLRW}$}{FLRW} Field Equations for Phantom EGB Theory}}
\textcolor{black}{In this appendix, we have shown from variational principle the derivation of the homogeneous field equations used in the main text. In our work and in this appendix, we have chosen:
$\mathcal L_m = \rho_m(t) = \rho_{m_0}\,a^{-3}(t)$}

\subsubsection{\textcolor{black}{Conventions and action}}

\textcolor{black}{Note that we have adopted the following conventions throughout:
\begin{itemize}
    \item Metric signature $(-,+,+,+)$.
    \item Units with $8\pi G = 1$ (so Einstein's equations read $G_{\mu\nu}=T_{\mu\nu}$).
    \item Spatially flat FLRW line element (cosmic time):
    \[
    ds^2 = -dt^2 + a^2(t)\,d\mathbf{x}^2,\qquad H=\frac{\dot a}{a}.
    \]
    \item Homogeneous scalar field: $\phi=\phi(t)$.
    \item Matter Lagrangian (explicit choice used here and in the main text ): $\mathcal L_m = +\rho_m(t)=\rho_{m_0}a^{-3}(t)$.
    \item Phantom kinetic term: the action contains a negative kinetic term for $\phi$ (this is the phantom choice).
\end{itemize}
}
\textcolor{black}{Along with the above conventions, the action we have considered in the paper is:
\begin{equation}\label{ac}
S=\int d^4x\sqrt{-g}\,\Big[\,R - f(\phi)\,G \;-\; \tfrac{1}{2}g^{\mu\nu}\nabla_\mu\phi\nabla_\nu\phi - V(\phi) - g(\phi)\,\mathcal L_m \,\Big],
\end{equation}
where $G$ is the Gauss–Bonnet scalar and given by:
\begin{equation}\label{GB}
 G \equiv R^2 - 4R_{\mu\nu}R^{\mu\nu} + R_{\mu\nu\rho\sigma}R^{\mu\nu\rho\sigma}.   
\end{equation}
We have considered the flat FLRW metric and hence:
\begin{equation}
 R = 6(\dot H + 2H^2), \qquad
G = 24H^2(\dot H + H^2), \qquad
\sqrt{-g}=a^3.   
\end{equation}}
\subsection{Variation with respect to the metric: Einstein equations}
We know that varying the action \eqref{ac} with respect to \(g^{\mu\nu}\) produces the modified Einstein equations in the form:
\begin{equation}\label{meq}
 G_{\mu\nu} \equiv R_{\mu\nu}-\tfrac{1}{2}Rg_{\mu\nu} = T^{\phi}_{\mu\nu} + T^{G}_{\mu\nu} + g(\phi)\,T^{(m)}_{\mu\nu},   
\end{equation}
where:
\begin{itemize}
    \item \(T^{(m)}_{\mu\nu} = -\dfrac{2}{\sqrt{-g}}\dfrac{\delta(\sqrt{-g}\,\mathcal L_m)}{\delta g^{\mu\nu}}\) is the matter stress tensor (for dust \(T^{(m)}_{00}=\rho_m,\;T^{(m)}_{ij}=0\)).
    \item \(T^{\phi}_{\mu\nu}\) is the scalar field stress tensor coming from the kinetic and potential pieces.
    \item \(T^{G}_{\mu\nu}\) is the effective stress tensor originating from the \(-f(\phi)G\) term.
\end{itemize}

For the homogeneous scalar \(\phi=\phi(t)\) the scalar stress tensor components give the standard phantom energy density and pressure:
\[
\rho_\phi \equiv T^{\phi}_{00} = -\tfrac{1}{2}\dot\phi^2 + V(\phi),\qquad
p_\phi \equiv T^{\phi}_{ii}/a^2 = -\tfrac{1}{2}\dot\phi^2 - V(\phi).
\]
(The negative kinetic contribution follows directly from the \(-\tfrac12 g^{\mu\nu}\nabla_\mu\phi\nabla_\nu\phi\) term in the action.)

The Gauss–Bonnet coupling contributes effective FLRW energy density and pressure pieces (standard results in the scalar–GB literature). One convenient form for these is:
\begin{align*}
\rho_G &= 24\,f_{,\phi}\,H^2\,\dot\phi,\\[4pt]
p_G &= -16(H^2+\dot H)H f_{,\phi}\dot\phi - 8H\big(\dot\phi^2 f_{,\phi\phi} + f_{,\phi}\ddot\phi\big).
\end{align*}
(These combinations appear when evaluating the $00$ and $ii$ components after variation; different algebraic rearrangements in the literature are equivalent up to total derivatives.)

Because the matter term in the action is \(-g(\phi)\mathcal L_m\), the metric variation places a factor \(+g(\phi)\) multiplying the usual matter stress tensor on the right-hand side; for dust this produces a \(+g(\phi)\rho_m\) contribution in the Friedmann equation.
Thus, when we consider all the contributions, the Friedmann (00) and acceleration (ii) equations become
\begin{equation}
 3H^2 = \rho_{\rm eff},\qquad 2\dot H + 3H^2 = -p_{\rm eff}~~~~~~~~~
with
~~~~~~~~~~\rho_{\rm eff} = \rho_\phi + \rho_G + g(\phi)\rho_m,
\qquad
p_{\rm eff} = p_\phi + p_G.   
\end{equation}
After substituting these expressions, we derive the field equations given in the main text as \eqref{F2} and \eqref{F3}:
\begin{equation}\label{frf}
3H^2 = -\tfrac{1}{2}\dot\phi^2 + V(\phi) + g(\phi)\,\rho_{m_0}a^{-3} + 24\,f_{,\phi}\,H^2\,\dot\phi,
\end{equation}
\begin{equation}\label{acf}
2\dot H + 3H^2 = \tfrac{1}{2}\dot\phi^2 + V(\phi) + 16(H^2+\dot H)H f_{,\phi}\dot\phi + 8H(\dot\phi^2 f_{,\phi\phi} + f_{,\phi}\ddot\phi)\Big.
\end{equation}

\subsection{Variation with respect to \texorpdfstring{\(\phi\)}: Klein–Gordon equation}
In order to derive the final equation, we vary the action, once again, with respect to $\phi$. we have listed the term-by-term contribution below:

\textcolor{black}{\begin{itemize}
    \item \textbf{Kinetic term:} variation of \(-\tfrac12\sqrt{-g}g^{\mu\nu}\partial_\mu\phi\partial_\nu\phi\) gives \(\sqrt{-g}\,(\ddot\phi + 3H\dot\phi)\).
    \item \textbf{Potential:} variation yields \(-\sqrt{-g}\,V_{,\phi}\).
    \item \textbf{GB coupling:} \(-\delta(\sqrt{-g}f(\phi)G)\) contributes \(-\sqrt{-g}\,f_{,\phi}G\) to the scalar equation of motion (the full metric-dependence also feeds into \(T^G_{\mu\nu}\), already accounted for above).
    \item \textbf{Matter coupling:} \(-\delta(\sqrt{-g}g(\phi)\mathcal L_m) = -\sqrt{-g}\,g_{,\phi}\mathcal L_m\). With the present choice \(\mathcal L_m=\rho_m\) this equals \(-\sqrt{-g}\,g_{,\phi}\rho_m\).
\end{itemize}
}
Therefore the KG equation (dividing by \(\sqrt{-g}\)) reads
\[
\underbrace{(\ddot\phi + 3H\dot\phi)}_{\text{kinetic/friction}} \;-\; V_{,\phi} \;-\; f_{,\phi}\,G \;-\; g_{,\phi}\,\rho_m \;=\; 0,
\]
or explicitly, substituting \(G=24H^2(\dot H + H^2)\) and \(\rho_m=\rho_{m_0}a^{-3}\):
\begin{equation}\label{eq:KG_final}
\boxed{\;\ddot\phi + 3H\dot\phi \;-\; V_{,\phi} \;-\; 24H^2(\dot H + H^2)f_{,\phi} \;-\; g_{,\phi}\,\rho_{m_0}a^{-3} \;=\;0\; }.
\end{equation}
Here, one may note that:
\begin{itemize}
    \item The kinetic piece appears with a \emph{positive} \((\ddot\phi+3H\dot\phi)\) in the equation because the action has a \emph{negative} kinetic term (\(-\tfrac12\cdots\)). This is equivalent to setting \(\epsilon=-1\) in the general \( (\epsilon/2)\,(\partial\phi)^2\) notation used elsewhere.
    \item The GB contribution is \(-f_{,\phi}G\) always (variation of \(-fG\) w.r.t.\ \(\phi\)).
    \item The matter source is \(-g_{,\phi}\mathcal L_m\); with \(\mathcal L_m=+\rho_m\) this is \(-g_{,\phi}\rho_m\). (If we had chosen \(\mathcal L_m=-\rho_m\) instead, the sign would flip which one may find in other various works in literature.)
\end{itemize}

Additionally, since matter is not conserved separately(due to non-minimal coupling), from
\(\nabla^\mu\big(T^\phi_{\mu\nu} + T^{G}_{\mu\nu} + g(\phi)T^{(m)}_{\mu\nu}\big)=0\) one obtains, for pressureless dust,
\[
\dot\rho_m + 3H\rho_m = -\frac{g_{,\phi}}{g(\phi)}\,\rho_m\,\dot\phi.
\]
This equation is consistent with the KG sourcing \(-g_{,\phi}\rho_m\): energy lost/gained by matter flows into/out of the scalar sector and GB-exchange channels, preserving total energy conservation.
Thus, the final system of equations we have derived and used in the main text as well is:

\begin{equation}
   \boxed{
\begin{aligned}
3H^2 &= -\tfrac{1}{2}\dot\phi^2 + V(\phi) + g(\phi)\,\rho_{m_0}a^{-3} + 24\,f_{,\phi}\,H^2\,\dot\phi \\[6pt]
2\dot H + 3H^2 &= \tfrac{1}{2}\dot\phi^2 + V(\phi) + 16(H^2+\dot H)H f_{,\phi}\dot\phi + 8H(\dot\phi^2 f_{,\phi\phi} + f_{,\phi}\ddot\phi) \\[6pt]
&\ddot\phi + 3H\dot\phi - V_{,\phi} - 24H^2(\dot H + H^2)f_{,\phi} - g_{,\phi}\,\rho_{m_0}a^{-3}= 0
\end{aligned}
}
\end{equation}


\vspace{0.5em}
\noindent\textbf{End of appendix.}

\section*{Declaration of generative AI and AI-assisted technologies in the writing process}

During the preparation of this work the author(s) used Grammarly and Quillbot in order to improve the language and correct the grammar. After using this tool/service, the author(s) reviewed and edited the content as needed and takes full responsibility for the content of the publication.

\section*{Acknowledgement}
The authors thankfully acknowledge the anonymous reviewer for valuable comments, which were immensely helpful for improving the manuscript.

\section*{Data Availability Statement}
The Type Ia supernova data used in this work come from the Pantheon+SH0ES compilation which is publicly available at \url{https://github.com/PantheonPlusSH0ES/DataRelease/tree/main/Pantheon%2B_Data} and   
the Planck 2018 $r$–$n_s$ confidence contours were obtained from \url{https://github.com/kdemirel/Planck-constraints-r-vs-ns-plot}.

\end{document}